\documentclass[%
 reprint,
superscriptaddress,
 amsmath,amssymb,
 aps,
floatfix,
]{revtex4-2}

\usepackage{graphicx, subcaption}% Include figure files
\graphicspath{ {./images3/} }
\usepackage{dcolumn}% Align table columns on decimal point
\usepackage{bm}% bold math
\usepackage{gensymb}
\usepackage{rotating}
\usepackage{booktabs}
\usepackage{xcolor}

\begin{document}

\preprint{APS/123-QED}

\title{Simulation of reversible molecular mechanical logic gates and circuits}

\author{Ian Seet}
\affiliation{ 
Physical and Theoretical Chemistry Laboratory, Department of Chemistry,University of Oxford, South Parks Road, Oxford OX1 3QZ, United Kingdom%\\This line break forced with \textbackslash\textbackslash
}%

\author{Thomas E. Ouldridge}
\affiliation{ 
Department of Bioengineering,
Imperial College London,
Royal School of Mines,
Exhibition Road,
London,
SW7 2AZ,
United Kingdom%\\This line break forced with \textbackslash\textbackslash
}%

\author{Jonathan P.K. Doye}
\affiliation{ 
Physical and Theoretical Chemistry Laboratory, Department of Chemistry,University of Oxford, South Parks Road, Oxford OX1 3QZ, United Kingdom%\\This line break forced with \textbackslash\textbackslash
}%

\date{\today}% It is always \today, today,

\begin{abstract}
Landauer's principle places a fundamental lower limit on the work required to perform a logically irreversible operation. Logically reversible gates provide a way to avoid these work costs, and also simplify the task of making the computation as a whole thermodynamically reversible. The inherent reversibility of mechanical logic gates would make them good candidates for the design of practical logically reversible computing systems if not for the relatively large size and mass of such systems. In this paper, we outline the design and simulation of reversible molecular mechanical logic gates that come close to the limits of thermodynamic reversibility even under the effects of thermal noise, and outline associated circuit components from which arbitrary combinatorial reversible circuits can be constructed and simulated. We demonstrate that isolated components can be operated in a thermodynamically reversible manner, and explore the complexities of combining components to implement more complex computations. Finally, we demonstrate a method to construct arbitrarily large reversible combinatorial circuits using multiple external controls and signal boosters with a working half-adder circuit.
\end{abstract}

\maketitle

%\tableofcontents

\section{\label{sec:intro}Introduction}
As the global power consumption of electronic devices rises in both absolute and relative terms, methods of reducing the per-operation energy cost of computational operations become increasingly important. It has been estimated that the switching energy of a standard CMOS transistor has a lower bound of 100\,$k_\mathrm{B}T$ \cite{Mam15}. Even if improvements in transistor technology allow this limit to be bypassed, a logically irreversible system will require a minimum of $k_\mathrm{B}T\ln{2}$ of work (or equivalent resources) spent for every bit erased \cite{Lan61, Fred82}. Performing such operations in a thermodynamically reversible manner, and recovering the work put in, is theoretically possible but impractical \cite{Brit21}.

Proposals for reversible electronic logic gates generally revolve around superconducting Josephson junction-based logic gates \cite{Sem07, Tak17}. Although capable of very high efficiencies, superconducting logic gates suffer from two weaknesses: firstly, superconductors require cryogenic temperatures to function; secondly, the use of inductors in superconducting logic gates substantially increases the difficulty of miniaturising such systems, with state-of-the-art superconducting logic circuitry having a much lower density than CMOS-based circuits \cite{Sem17}.

Logically reversible circuits can also be constructed from traditional molecular logic systems as used in DNA computing \cite{Gen11, Tao14}. Such systems generally rely on diffusion to transmit information and bond-breaking processes to execute logical operations. Although it is certainly possible to create practical logically reversible gates via this approach, DNA logic gates tend to be very slow precisely due to their reliance on diffusion and bond-breaking processes.

Mechanical logic gates are another possibility for implementing reversible logic. Although macroscale mechanical logic is among the oldest computing paradigms in existence \cite{Free06, Brom98}, relatively little work has been done on micro- and nanoscale mechanical logic. Several such designs have been proposed \cite{Yas21, Mer93, Wen11, Mer18, Song19}. These designs, though often heavily optimised to reduce inter-component friction, possess kinetic energies much higher than $k_\mathrm{B}T$ when operating at practical clock speeds, even for a single gate. As all the designs considered require constant starting and stopping of moving components during particular phases of the clock cycle, failing to recover this energy will result in efficiency losses far greater than $k_\mathrm{B}T$.

Attempting to recover the kinetic energy with nanoscale springs is unlikely to be very efficient. The Q factor is the ratio between the energy contained within a damped harmonic oscillator at the start of a cycle and the energy lost after one radian of oscillation, and is inversely proportional to the damping ratio \cite{sly07}; thus, it functions as an upper bound on the efficiency of kinetic energy scavenging. The Q factor of an atomic force microscopy cantilever in ultra-high vacuum is of the order $10^{3}$ to $10^{4}$ in the 10--100\,MHz range \cite{Kaw02}. The Q factor is much lower at regular atmospheric pressure, being of the order $10^{2}$ in the 5--50\,kHz range \cite{Per12}. These values, however, cannot be treated as indicative of the energy recovery potential of a realistic mechanical logic device since the logic gates themselves are neither perfectly rigid nor perfectly elastic, and will probably have significantly lower Q factors than purpose-built cantilevers. Furthermore, Q factors generally decrease with increasing frequency; coupled with the quadratic increase in kinetic energy with clock speed in mechanical systems, this would result in a super-quadratic increase of energy dissipation per clock cycle with respect to clock speed. In addition, the restoring force imposed by the springs would likely increase friction between components, leading to a further decrease in efficiency.

One method to avoid the losses associated with kinetic energy recovery is to use molecular mechanical logic gates, here defined as systems  with kinetic energy that does not significantly exceed that predicted by equipartition even at high clock speeds.  In order to lower the kinetic energy, a major reduction in size is necessary, greatly increasing the difficulty of avoiding inter-component drag (analogous to friction in macroscale systems) as components will often impinge on the van der Waals radius of others. In addition, the fact that the entire system must be close to thermal equilibrium precludes any reliance on momentum to ensure the correct positioning of moving components. Therefore, every step of every transition must be a local potential energy minimum with a sufficiently high free-energy barrier to prevent unwanted transitions into erroneous states.

Despite these additional constraints, molecular mechanical logic gates do possess some major advantages over their larger counterparts. Other than the obvious advantage of being smaller and more compact, they would in addition be capable of higher efficiencies at high clock speeds due to inter-component drag and/or viscous drag dominating over the imperfect conversion of kinetic energy as the primary mode of energy loss. These gates would also perform much more efficiently under normal atmospheric pressure or liquid media due to experiencing far less viscous drag due to their smaller size, though it would be unlikely they could achieve the same efficiencies as larger, low-friction gates at very low clock speeds and ultra-high vacuum.

At present, the techniques of chemical synthesis do not allow for the efficient construction of molecular mechanical logic gates. However, semi-realistic coarse-grained simulations of molecular mechanical circuits would nonetheless allow exploration of the thermodynamic limits of computation in such systems. Although imperfect, such models would also provide significant advantages over purely theoretical reasoning \cite{Brit21, Kol20}. Indeed, the majority of work on the thermodynamics of computing hitherto has either considered extremely abstract models \cite{Kol20} or very simple operations with explicit systems, such as bit erasure \cite{Kar2020} or copying of individual bits \cite{Des17}, limiting understanding of the physical constraints on more complex computational systems.

In this paper, we demonstrate the results from simulations of an efficient novel molecular mechanical NAND gate design and its associated circuit components, with the capability of generating combinatorial circuits with arbitrary relative positioning in three dimensions. This model allows us to explore the consequences of building larger circuits with multiple gates. We show that we can chain such NAND gates into a combinatorial circuit that operates with low error rate and close to the thermodynamic limits of efficiency with a working half-adder circuit. We also demonstrate the unsuitability of directly chaining gates together when a low error rate is desired.

\section{\label{sec:methods}Methods of Simulation}
A coarse-grained rigid-body simulator and basic modelling components were used to simulate the systems studied in this paper. The large, medium and small coloured beads in the figures and animations represent three distinct particle types, corresponding approximately to masses of 120, 41 and 14 amu respectively. The beads are not meant to represent any particular organic moiety, as it is not our intention to create an accurate model of a specific structural motif.

Intermolecular forces between each pair of coloured beads were modelled with a modified Weeks-Chandler-Anderson (WCA) potential (Eq.\ \ref{eq:wca}). The WCA potential is itself a modified Lennard-Jones potential which is increased by $\epsilon$ and truncated at the Lennard-Jones potential minimum \cite{Wee71}. This modification ensures that there is no long-range attractive term, which allows the cut-off radius of the Verlet list to be significantly reduced, speeding up calculations.
\begin{eqnarray}
V_\mathrm{WCA}(r) =
\begin{cases} 
  4\epsilon[(\frac{\sigma}{r - d})^{12} - (\frac{\sigma}{r - d})^6] + \epsilon, r - d \leq 2^{\frac{1}{6}}\sigma\\
  0,\qquad\qquad\qquad\qquad\qquad r - d > 2^{\frac{1}{6}}\sigma\mathrm{,}
\end{cases}
\label{eq:wca}
\end{eqnarray}
where $r$ is the distance between the two particles, $\epsilon$ and $\sigma$ are the potential depth and van der Waals radius constants from the standard Lennard-Jones potential while $d$ represents a step-off distance to account for the fact that the beads are meant to represent a cluster of atoms rather than a single atom.

Physically reasonable values of $\epsilon$ and $\sigma$ were obtained by fitting to appropriate intermolecular potentials. For the large beads, these values were obtained by plotting the Lennard-Jones potential of two adamantane molecules from a distance of 0.5 to 2.0\,nm using the Generalized Amber Force Field (GAFF) parameters for carbon and hydrogen \cite{amber}, while $d$ was determined by least-squares fitting. Values of $\epsilon = 3.33$, $\sigma = 0.2$\,nm and $d = 0.15$\,nm were obtained; $\epsilon$ values are given in units of $k_\mathrm{B}T$ at 298\,K. The step-off distance ensured a more accurate fit to the plotted values than would otherwise be possible with the standard Lennard-Jones potential calculated using the distance between the centres of two beads. For the small beads, values of $\epsilon = 0.2$, $\sigma = 0.19$\,nm and $d = 0$\,nm were obtained directly from the GAFF parameters for the methyl group. The radius parameters $\sigma = 0.14$\,nm and $d = 0.11$\,nm for the medium-sized beads were obtained by scaling down the constants for the large beads by assuming a cubic relation between van der Waals radius and mass, while the value of $\epsilon = 0.8$ was found by assuming that the mass of a bead $m$ is approximately related to $\epsilon$ via a simple polynomial equation $\epsilon = Am^n$, where $A$ and $n$ are constants to be solved for using the known values of $m$ and $\epsilon$ of the small and large beads.

To calculate the values of $\sigma$, $d$ and $\epsilon$ between two different bead types, the expressions
\begin{eqnarray}
    \sigma_{ij} = (\sigma_i + \sigma_j)/2
\label{eq:ljsigma}
\end{eqnarray}
\begin{eqnarray}
    d_{ij} = (d_i + d_j)/2
\label{eq:ljd}
\end{eqnarray}
\begin{eqnarray}
    \epsilon_{ij} = \sqrt{\epsilon_i\epsilon_j}
\label{eq:ljepsilon}
\end{eqnarray}
were used to find the mixed parameters $\sigma_{ij}$, $d_{ij}$ and $\epsilon_{ij}$ for each bead type $i$ and $j$.

A quaternion-based Langevin thermostat \cite{Dav14} was used to maintain the temperature of the canonical ensemble and generate Brownian motion. The motion of particles is governed by the Langevin equation of both translational and rotational coordinates and momenta. The Hamiltonian of the system is $H(\mathbf{r}^N, \mathbf{p}^N, \mathbf{q}^N, \boldsymbol{\Pi}^N)$ where $\mathbf{r}$ and $\mathbf{p}$ are the positions and momenta of each of the $N$ rigid bodies in the system, while $\mathbf{q}$ and $\boldsymbol{\Pi}$ are the orientation and angular momenta of each rigid body expressed in terms of quaternions. Using this Hamiltonian formalism, the translational and rotational Langevin equations can be expressed as
\begin{eqnarray}
\dot{\mathbf{p}_i} = -\frac{{\partial H}}{{\partial \mathbf{r}_i}} -{\gamma_i}\mathbf{p}_i + \sqrt{\frac{2m_i{\gamma_i}}{\beta}}\mathbf{w}_i(t)\mathrm{,}
\label{eq:transLang}
\end{eqnarray}
\begin{eqnarray}
\dot{\boldsymbol{\Pi}_i} = -\frac{{\partial H}}{{\partial \mathbf{q}_i}} -{\Gamma_i}\mathbf{G}(\mathbf{q}_i, \boldsymbol{\Pi}_i) + \sqrt{\frac{2M_i{\Gamma_i}}{\beta}}\mathbf{W}_i(t)\mathrm{,}
\label{eq:rotLang}
\end{eqnarray}
\begin{eqnarray}
M_i = \frac{4}{\mathrm{Tr}(I_i^{-1})}
\label{eq:moiLang}
\end{eqnarray}
where $\beta$ is $1/k_{B}T$, $m_i$ is the mass, $\mathrm{Tr}(I_i^{-1})$ is the trace of the inverse of the moment of inertia tensor in the principal axis frame, and ${\gamma_i}$ and ${\Gamma_i}$ are the translational and rotational damping constants of the $i^\mathrm{th}$ rigid body. $\mathbf{G}(\mathbf{q}_i, \boldsymbol{\Pi}_i)$ is a four-dimensional vector that couples the angular momentum quaternion with the orientation quaternion, while $\mathbf{w}_i(t)$ is a three-dimensional Gaussian random variable and $\mathbf{W}_i(t)$ a four-dimensional quaternion Gaussian random variable representing noise due to Brownian motion.  To solve the aforementioned Langevin equations numerically we used Langevin Integrator `C' from the work of Davidchak \textit{et al.}\ \cite{Dav14}. Importantly, this approach to the dynamics is thermodynamically self-consistent, i.e.\ it is capable of reproducing the expected Maxwell-Boltzmann distribution of kinetic energies and the equipartition theorem for a system in thermal equilibrium.

The translational and rotational damping constants for single beads were calculated using the equations for Stokes' drag of a spherical object:
\begin{eqnarray}
{\gamma} = 6{\pi}{\mu}(\sigma + d)/m
\label{eq:transStokes}
\end{eqnarray}
\begin{eqnarray}
{\Gamma} = 8{\pi}{\mu}(\sigma + d)^3\mathrm{Tr}(I^{-1})
\label{eq:rotStokes}
\end{eqnarray}
where $\mu$ is the viscosity of the solvent. For non-spherical agglomerations of multiple beads, the value of $\sigma + d$ was approximated using the radius for a solid sphere of identical mass and density.The dynamic viscosity of air at 298\,K ($1.81 \times 10^{-5}$ kg$\,\mathrm{m}^{-1}\mathrm{s}^{-1}$) was used in all simulations. At this viscosity, the system is overdamped within the range of clock speeds tested, and the kinetic energy of a single logic gate and its associated drivers does not exceed 0.3\,$k_\mathrm{B}T$ beyond what would be expected from equipartition even at the highest clock speed tested (2.87\,GHz).  The damping term applied is only an approximation of the true drag experienced by the rigid bodies as it neglects the asphericity of the bodies and does not take into account their hydrodynamic interactions. However, a highly accurate model of drag is neither necessary nor useful given that the system is designed to be a coarse-grained representation of large rigid molecules and not a precise model of a specific molecular system. 

The simple harmonic potentials $V_{\mathrm{bond}} = k_{\mathrm{bond}}(r - r_0)^2/2$ and $V_{\mathrm{angle}} = k_{\mathrm{angle}}({\theta} - {\theta}_0)^2/2$ were used to constrain the relative distances and angles between the different components of the logic gate. The bond harmonic constants $k_\mathrm{bond}$ were all set to 214\,N$\,\mathrm{m}^{-1}$, approximately half the strength of a $sp^3$ carbon-carbon single bond from GAFF, while the angular harmonic constants $k_\mathrm{angle}$ were set to 520\,kJ\,$\mathrm{mol}^{-1}\mathrm{rad}^{-2}$, approximately equal to the bending potential of a chain of three $sp^3$ carbon atoms; however, it should be noted that as there are multiple angular potentials centered on a realistic carbon atom (typically two to three depending on the hybridization state of the central atom), the angular potential we have used would be weaker than the actual angular potential felt by two rigid $sp^3$ carbon-based structures bonded together. In both cases, the bond and angular potentials are substantially weaker than would be realistic; this attenuated potential compensates for the unrealistic stiffness of the rigid bodies that comprise the bulk of the system.

Two types of bonds were used to connect the rigid body systems. Both used the same bond harmonic potential, but differ in the angular potentials used to restrain the relative positions of their parent rigid bodies. The first type, coloured silver in all figures, represents a bonding interaction between two particles $A$ and $B$, and also implies two sets of angular potentials $\measuredangle{XAB}$ and $\measuredangle{ABY}$ where $X$ is an arbitrary position on the rigid body to which particle $A$ belongs that is fixed with nonzero distance relative to particle $A$, and $Y$ a similar position on the rigid body to which particle $B$ belongs. The second type, coloured yellow, represents only a bonding interaction between $A$ and $B$, without any angular potential.

An important part of our approach is that the external control used to drive the computations should be simple \cite{Brit21, Brit19}. Two reasons underlie this philosophy; firstly, it is common within the field of stochastic thermodynamics to assume that an arbitrary time-varying potential can be applied to the system's coordinates with essentially no external cost. However, it is unclear how efficient transfer of work to and from the computational system can actually occur in this setting. Secondly, the external costs of even a simple control can only be neglected if the system being controlled is relatively large \cite{Brit21}. Any control externally applied must therefore be applicable in parallel to many systems, such that the cost of application is amortised over all systems being driven in parallel.

Therefore, our systems are driven via one or more external dipoles that rotate at constant angular velocity until a change of direction is needed, at which point the angular velocity is negated. The dipole potential takes the form
\begin{eqnarray}
V_{\mathrm{dipole}} = k_{\mathrm{dip}}\mathbf{\hat{d}_{\mathrm{ext}}}\cdot\mathbf{\hat{d}_{\mathrm{int}}}
\label{eq:dipole}
\end{eqnarray}
where $\mathbf{\hat{d}_{\mathrm{ext}}}$ and $\mathbf{\hat{d}_{\mathrm{int}}}$ are the direction vectors of the external and internal system dipoles, respectively, and $k_{\mathrm{dip}} = 496\,\mathrm{kJ/mol}$ for all dipoles.  This interaction provides a simple generic means of rotary mechanical transmission that need not necessarily be implemented using electromagnetic dipoles. In keeping with this simplified approach we do not specify the position of the external dipole or include any distance dependence in the interaction. Each internal dipole interacts with the external dipole independently, and its phase relative to other internal dipoles within the system can be controlled.

Such a simple control mechanism that requires no feedback from the computational state of the machine makes the efficient transfer of work to and from the computational system more plausible. Moreover, a control of this kind allows, at least in principle, for a single external protocol that rotates multiple dipoles within one or more circuits together, provided that they maintain a fixed relative phase, thus allowing the external cost of the control to be neglected.  More sophisticated protocols with variations in relative phase or frequency could be physically implemented while maintaining only a single external dipole via the use of reduction gearing.   

The following expression \cite{Sek99} was used to calculate the work done on the gate by the driving force of a single dipole:
\begin{eqnarray}
E_{\mathrm{dipole}}(t, {\Delta}t) = k_{\mathrm{dip}}(\mathbf{\hat{d}_{\mathrm{ext}}}(t + {\Delta}t) - \mathbf{\hat{d}_{\mathrm{ext}}}(t)) \cdot\mathbf{\hat{d}_{\mathrm{int}}}
\label{eq:work}
\end{eqnarray}
where $\mathbf{\hat{d}_{\mathrm{ext}}}(t)$ is the direction vector of the external dipole at time $t$, ${\Delta}t$ is the length of one timestep, and $E_{\mathrm{dipole}}$ the work done by the external dipole on a given internal dipole across the timestep ${\Delta}t$.  All simulations were run with a timestep of 3.40\,fs.

\section{\label{sec:cc}Circuit Components}
In order to construct an arbitrary combinatorial circuit, we require: at least one type of universal Boolean logic gate (typically a NAND gate); a mechanism for connecting the input of one gate to the output of another; and a device capable of converting signals from the clock into a means of selectively transducing information between connected gates (henceforth referred to as a driver). As we are not demonstrating sequential logic in this paper, it is unnecessary to demonstrate a means of storing information beyond a single clock cycle. The driver-bit system in Section \ref{sec:landauer} was designed only as a means of verifying the thermostat's capability of replicating Landauer's principle rather than a practical means of information storage. We now introduce these basic composable components in detail.
\subsection{\label{ssec:driver}The Driver}
\begin{figure*}
	\includegraphics[width=\linewidth]{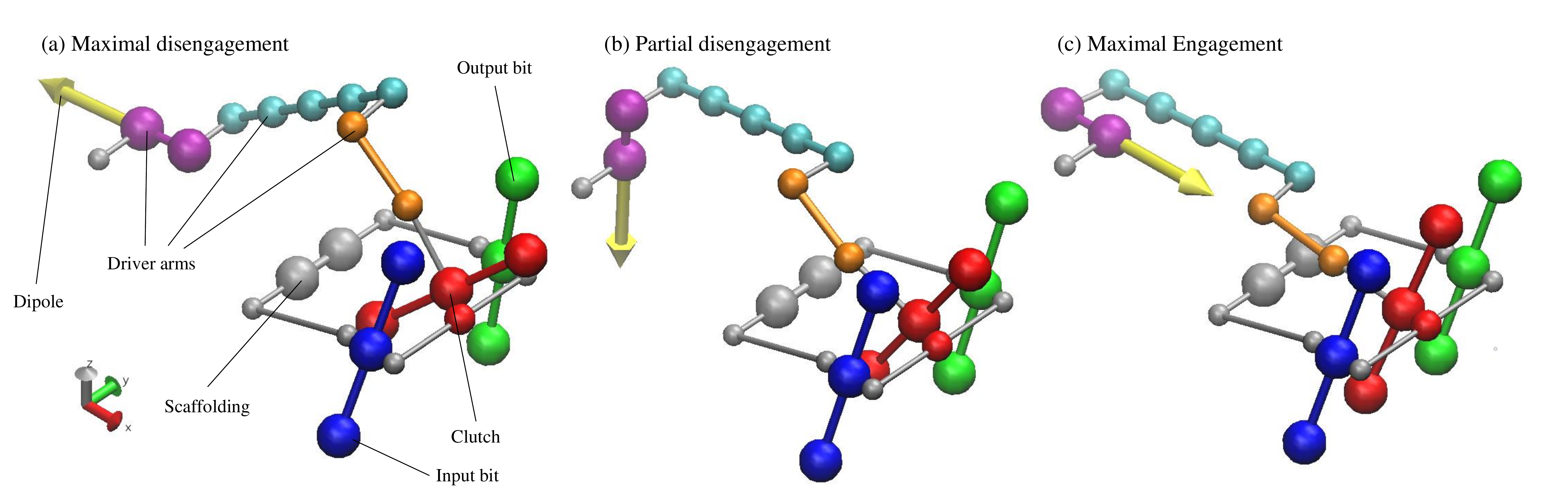}
	\caption{The driver system used to selectively transduce information depending on the phase of the clock. From left to right, these diagrams depict a right-handed driver in states of increasing degrees of engagement. Supplemental Video 1 provides an animation of this process.}
	\label{fig:driverR}
\end{figure*}

\begin{figure}
	\includegraphics[width=\linewidth]{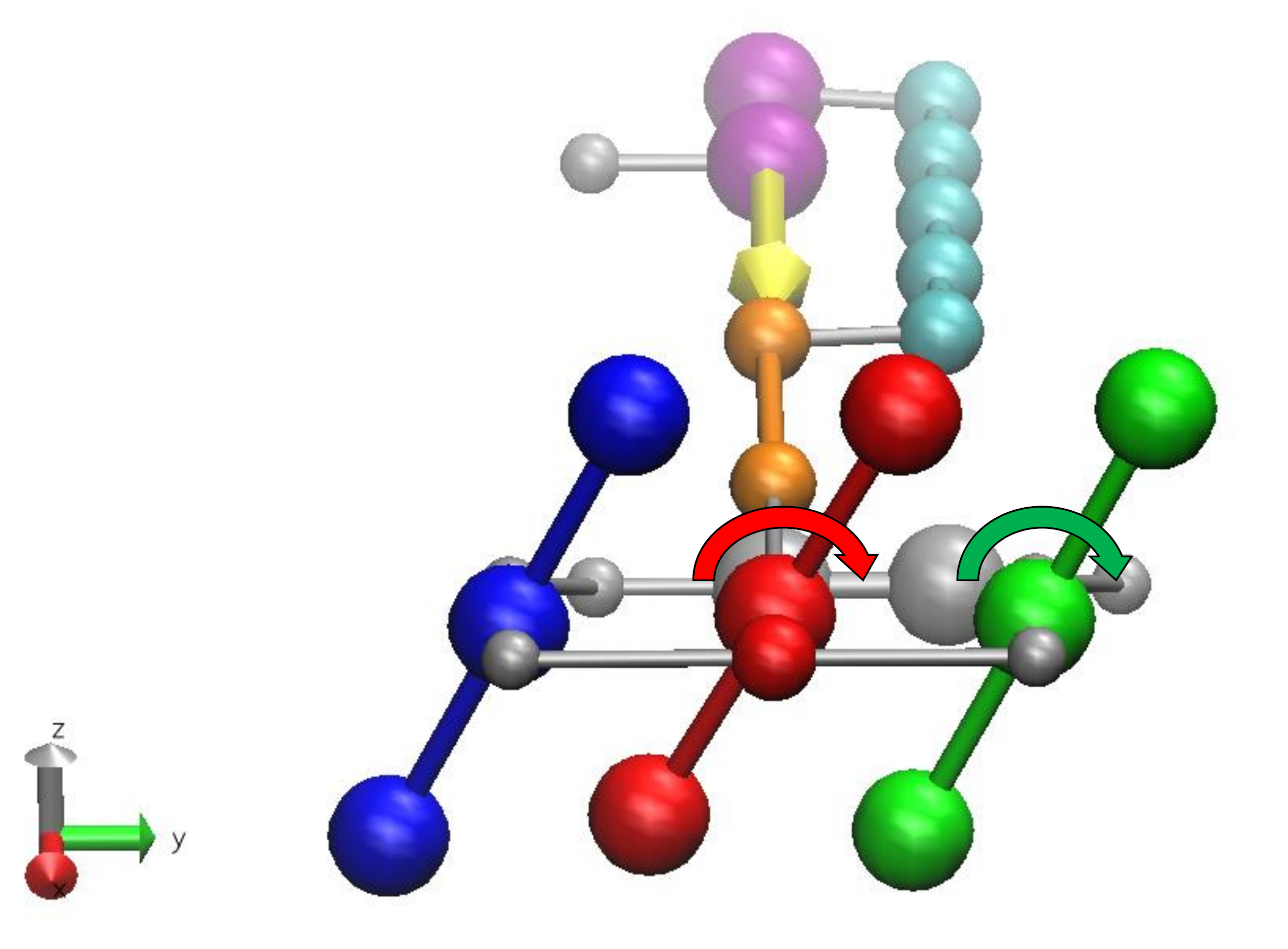}
	\caption{The driver system at maximal engagement visualised along the $x$-axis. The coloured arrows illustrate the ability of the clutch and output bit to rotate about the $x$-axis to match the state of the input bit. Note that the clutch is capable of rotating along two axes ($x$ and $y$).}
	\label{fig:driverR2}
\end{figure}

The driver is responsible for converting the mechanically-driven clock signal into a means of transducing information between a fixed input bit and a gate or between two gates via a clutch whose engagement is a function of the phase of the clock. Figs.\ \ref{fig:driverR} and \ref{fig:driverR2} illustrate the driver design as well as the general colouring scheme used in the circuits in this paper. Each rigid body is represented by objects of one colour; coloured bonds show that the particles they connect belong to the same rigid body. The differently-sized spheres represent the three different particle types used. As previously mentioned in Section \ref{sec:methods}, silver-coloured bonds represent bonding interactions with an angular component between rigid bodies, while yellow-coloured bonds in Fig.\ \ref{fig:wingedNAND} represent purely distance-based interactions. The silver beads represent a scaffold to which the rigid bodies are constrained.  Spheres directly bound to the scaffold are constrained by a harmonic potential centered on their initial position instead of being bonded directly to the scaffold.  The yellow arrow indicates an internal dipole that interacts with the clock-driven external dipole via the external dipole potential mentioned previously. A similar colouring scheme is used for diagrams throughout the paper.

The driver arms function in a manner similar to a piston rod. The external dipole rotates in the plane henceforth arbitrarily defined as the $xz$-plane with the $x$-axis corresponding to the alignment of the dipole at a state of minimal/maximal engagement. When the external dipole rotates, it forces the internal dipole to follow synchronously, moving the
 purple arm  of the driver into the position of greatest distance between it and the clutch (see Fig.\ \ref{fig:driverR}). The cyan and orange segments are pulled into a position such that they lie parallel to the purple arm, bringing the red clutch into the $yz$-plane  where it can transfer via steric interactions the state of the blue input bit to the green output bit.

When the  purple  arm is rotated to the position of minimal distance from the clutch, it forces the orange arm to take on a pitch of roughly {45\textdegree}  from the plane of the scaffold, forcing the clutch away and decoupling the output and input bits from each other (Fig.\ \ref{fig:driverR}a). In addition to providing a rigid attachment point, the scaffolding also prevents the driver arms from rotating below the plane defined by the scaffold, eliminating any possibility of the driver taking two different states, a scenario that complicates reversibility simulations.

The blue input bit is restrained to an angle of 45\textdegree\ from the $z$-axis by a dipole potential identical in strength to the one used for the interaction between the internal and external dipoles (Eq.\ \ref{eq:dipole}); bits with a positive slope in the $yz$-plane (pointing away from the viewer) are arbitrarily defined as `1' and bits with a negative slope as `0'. In Fig.\ \ref{fig:driverR}, all input bits have been set to `1'. As a circuit component, the driver does not always connect a restrained input bit to an output bit, but can also be inserted between the output bit of one gate and the input bit of another, allowing the selective transduction of information based on the clock cycle.

Due to the geometry of other circuit components, two distinct driver types are used throughout this paper, one arbitrarily designated as a right-handed driver (Fig.\ \ref{fig:driverR}) and the other a left-handed driver. Although the rigid bodies of the two driver types are mirror images of each other, the interaction potentials are not, as the internal dipole for each driver points in the same direction. Therefore, both drivers rotate in the same direction with the same phase.
\subsection{\label{ssec:nand}The NAND gate}
\begin{figure*}
	\includegraphics[width=\linewidth]{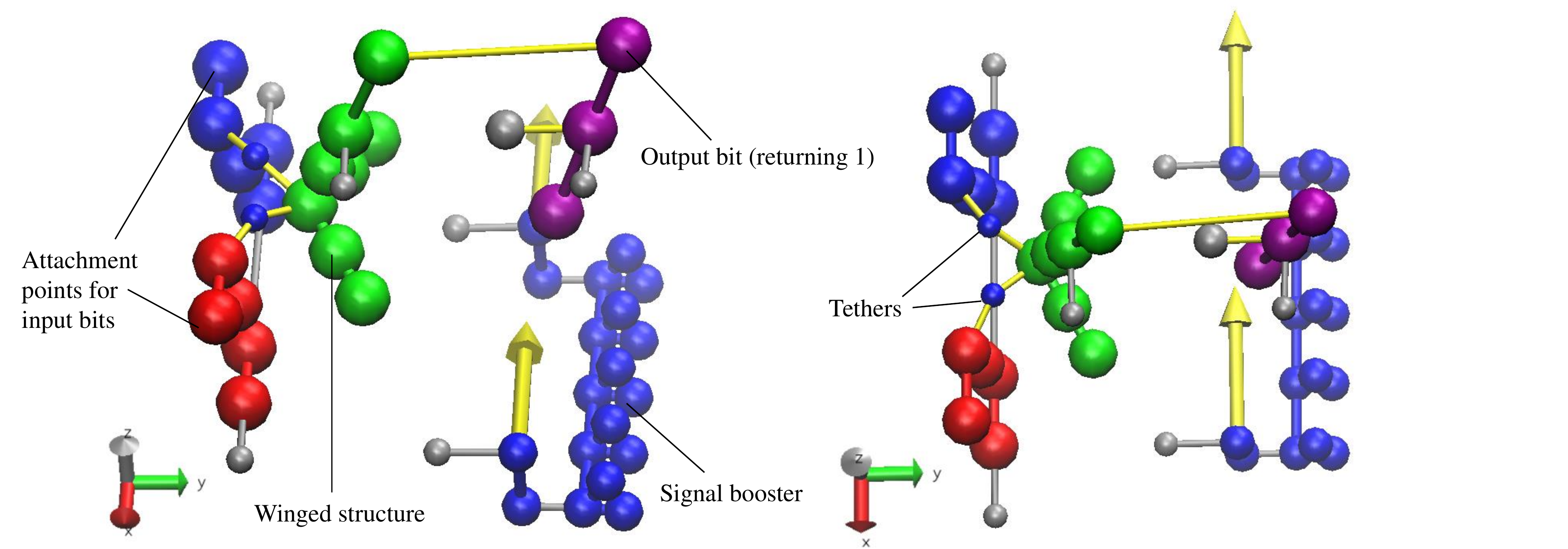}
	\caption{The winged NAND gate, which constitutes the basic universal Boolean gate used in circuit construction, viewed from two different orientations. Note that the winged structure can rotate freely about the $x$- and $z$-axes. The orientation of the winged structure is that for when the inputs are both zero.
	\label{fig:wingedNAND}
}
\end{figure*}
\begin{figure*}
	\includegraphics[width=\linewidth]{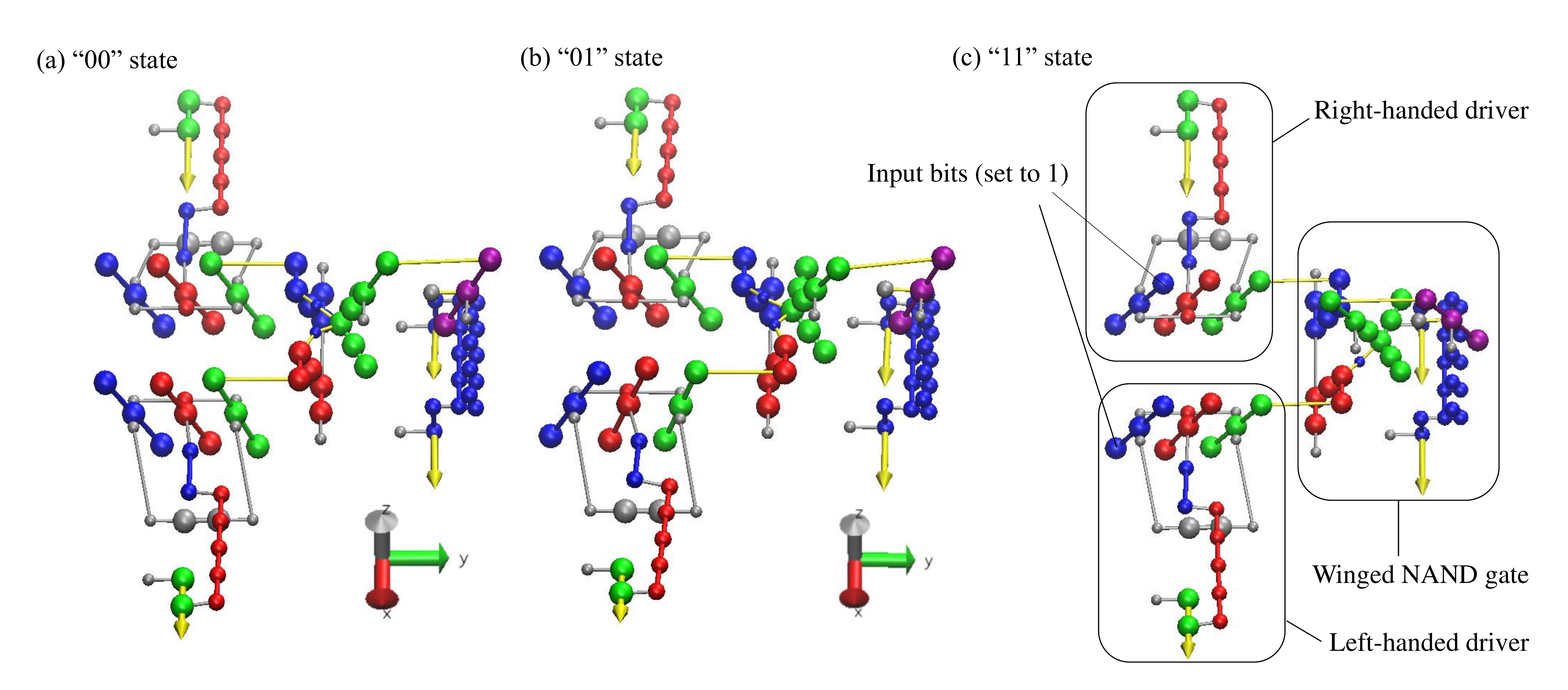}
	\caption{The winged NAND gate connected to two engaged drivers with inputs (a) `00', (b) `01' and (c) `11', illustrating how the states of the inputs affect the orientation of the winged structure and hence the output.
	\label{fig:NANDgallery}
}
\end{figure*}

Any combinatorial Boolean logic gate may be constructed from NAND gates alone. One of the most intuitive methods of designing a molecular mechanical NAND gate is to conceptualise it as a two-bit majority-voter gate. The $N$-bit majority voter gate accepts $N$ inputs and returns 1 if a majority of the $N$ inputs are set to 1; otherwise, it returns 0 \cite{nash17}; therefore, a two-bit majority voter gate is synonymous with an AND gate. NAND gates may in turn be trivially constructed from AND gates by inverting the output bit. A two-bit mechanical majority voter gate may be implemented in a fairly simple manner by using a spring to restrain the gate to a default position corresponding to the `0' state. This spring is countered by an opposing force that is applied by the input bits when shifted into the `1' state; however, this force is only sufficient to overcome the original restraining force when both input bits are shifted into the `1' state.

Despite its simplicity, the two-bit majority voter-based NAND gate does bear the disadvantage that its four possible states (`00', `01', `10' and `11') have significantly different energies due to the strain caused by the force exerted by the input bits acting against the restraining spring. While it may be theoretically possible to offset this energy imbalance by the addition of more springs, the structure would inevitably contain significant internal strain.

To avoid these potential problems, the molecular mechanical NAND gate outlined in Figs.\ \ref{fig:wingedNAND} and \ref{fig:NANDgallery} was designed. The output bit of the gate is connected to a freely-rotating pair of prong- or wing-like structures, hence the term ``winged'' NAND gate. The tip of the output bit is tethered to the tips of the input bits via the winged structure. When the input bits are fixed to the states `00', the  winged structure  is pulled by the input bits into the `1' configuration,  and the output bit with it. When the input bits are set to `01' or `10', the  winged structure  is simultaneously pulled forward by the `0'-set bit while being pushed forward by the `1'-set bit; however, the ability of the wings to freely rotate about the axis of the  winged structure  allows it to be pulled forward by the `0'-set input bit. When the input bits are set to `11', this rotation is not possible and the output bit is forced into the `0' configuration (Fig.\ \ref{fig:NANDgallery}). This design has the benefit of not intrinsically relying on internal strain to achieve the desired output, though it is acknowledged that practically designing a winged NAND gate with minimal strain is not trivial.

As the angular deflection of the gate output induced by the gate inputs is smaller than the deflection of the input bits, a signal booster is required to increase the deflection of the output bit. The signal booster is a wedge-shaped block that is driven by the external dipoles and remains in phase with the gate drivers. When the drivers are maximally engaged, it forces the output bit away from the $y$-axis, thereby boosting the signal.
\subsection{\label{ssec:switch}The Switch}
\begin{figure*}
	\includegraphics[width=\linewidth]{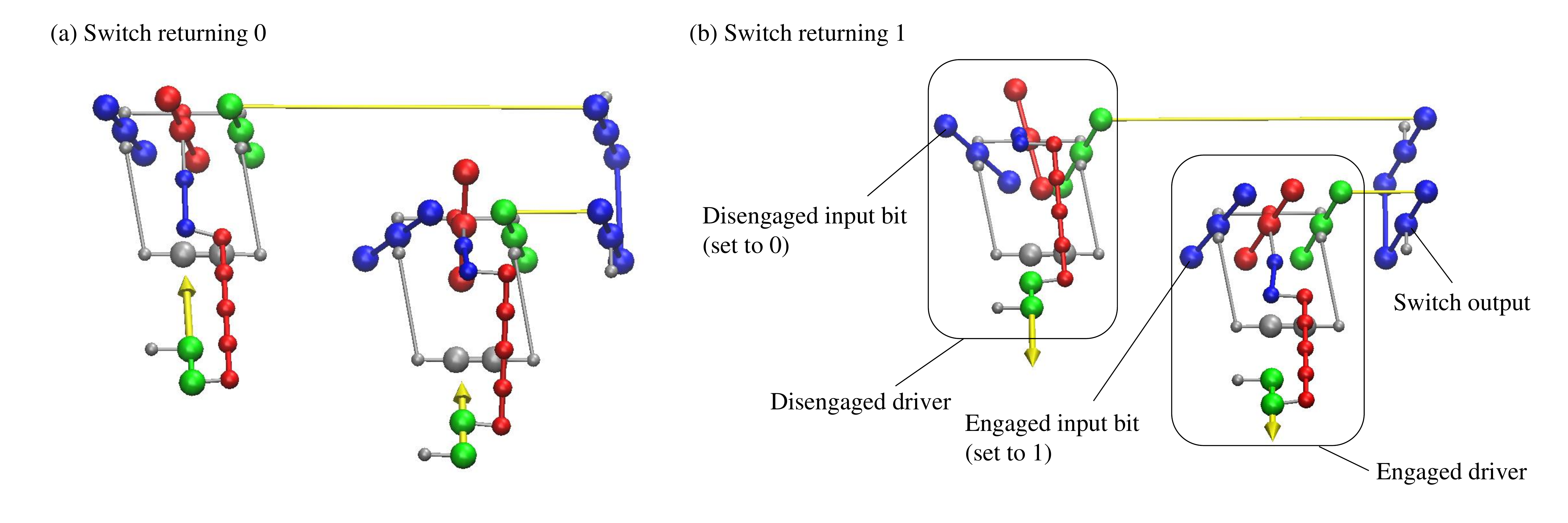}
	\caption{The switch design used to selectively transmit information from a given bit depending on the phase of the external dipole. The switch in (a) is currently in the `0' state and will transition to the `1' state if the external dipole advances by 180\textdegree; the reverse is true for the switch in (b).}
	\label{fig:switchGallery}
\end{figure*}
The switch provides a means of selectively relaying information to an output from two different input bits via a change in the phase of the external dipole without relying on a multiplexer. It consists of two drivers with a phase offset of 180\textdegree\ such that the internal dipoles of the two drivers point in opposite directions, tethered to a common switch output. By altering the phase of the external dipole, one of the input bits of the drivers can be selectively expressed at the output (Fig.\ \ref{fig:switchGallery}).

By connecting two pairs of switches to the NAND gate, the inputs of the NAND gate can be selectively varied with the phase of the external dipole (and hence, the clock). This allows for transitions between arbitrary pairs of states.
\subsection{\label{ssec:circon}Circuit construction}
\begin{figure}
	\includegraphics[width=\linewidth]{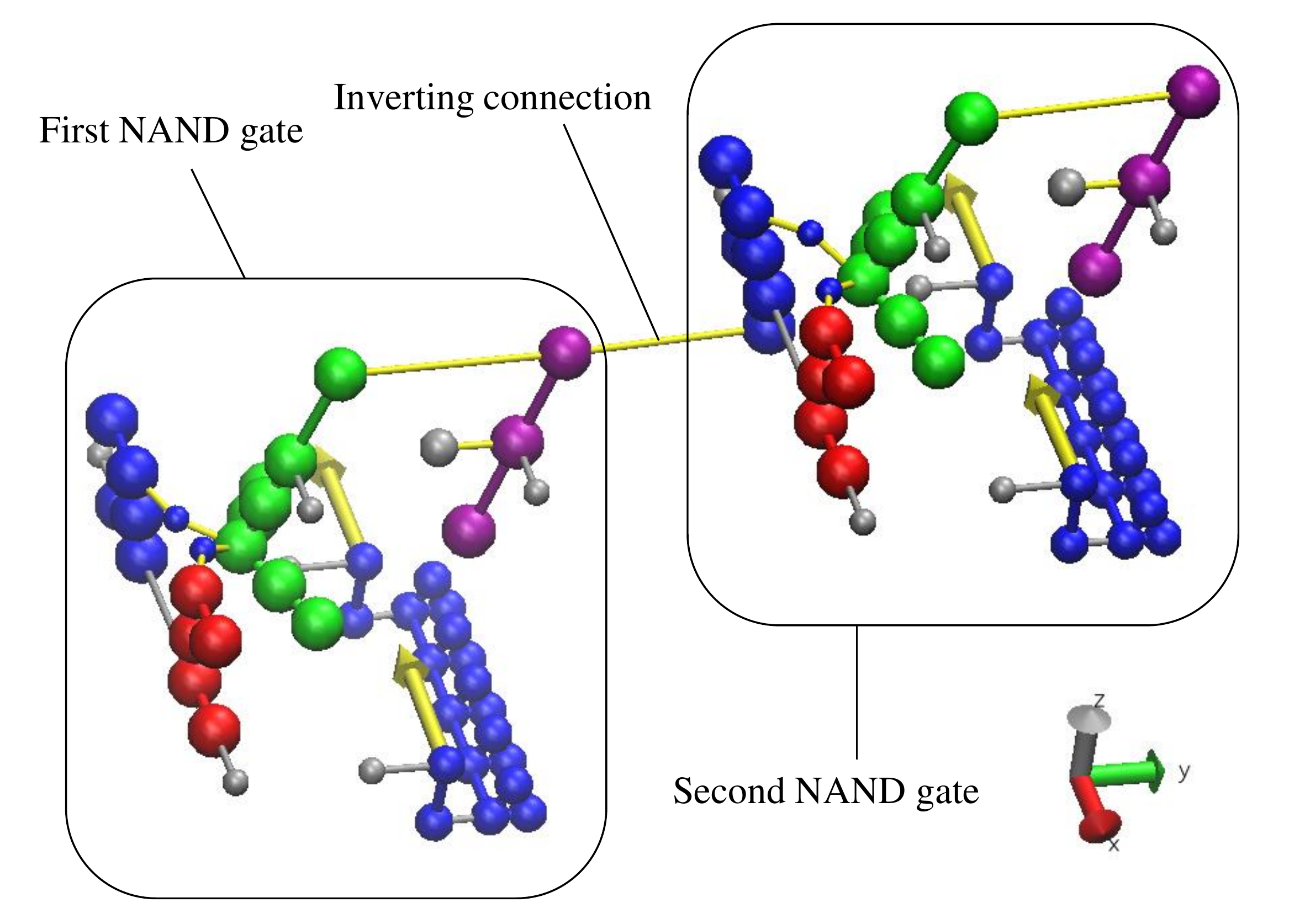}
	\caption{The output of one NAND gate fed into the input of another via an inverting connection.}
	\label{fig:inverter}
\end{figure}

Several methods of chaining individual logic gates into useful combinatorial circuits are possible. The simplest method is to directly tether the output of one gate to the input of the other. Direct tethering can also be used to fulfil the role of an inverter. Although the inverter can be constructed from a NAND gate by feeding the inverter input into both inputs of the NAND gate, this is generally an inefficient use of resources. It is more efficient to implement an inverter within this particular molecular mechanical logic system by tethering the output of one gate to another in a manner such that an increase in the slope of the output bit results in a decrease in the slope of the input bit, and vice versa (Fig.\ \ref{fig:inverter}).

\begin{figure}
	\includegraphics[width=\linewidth]{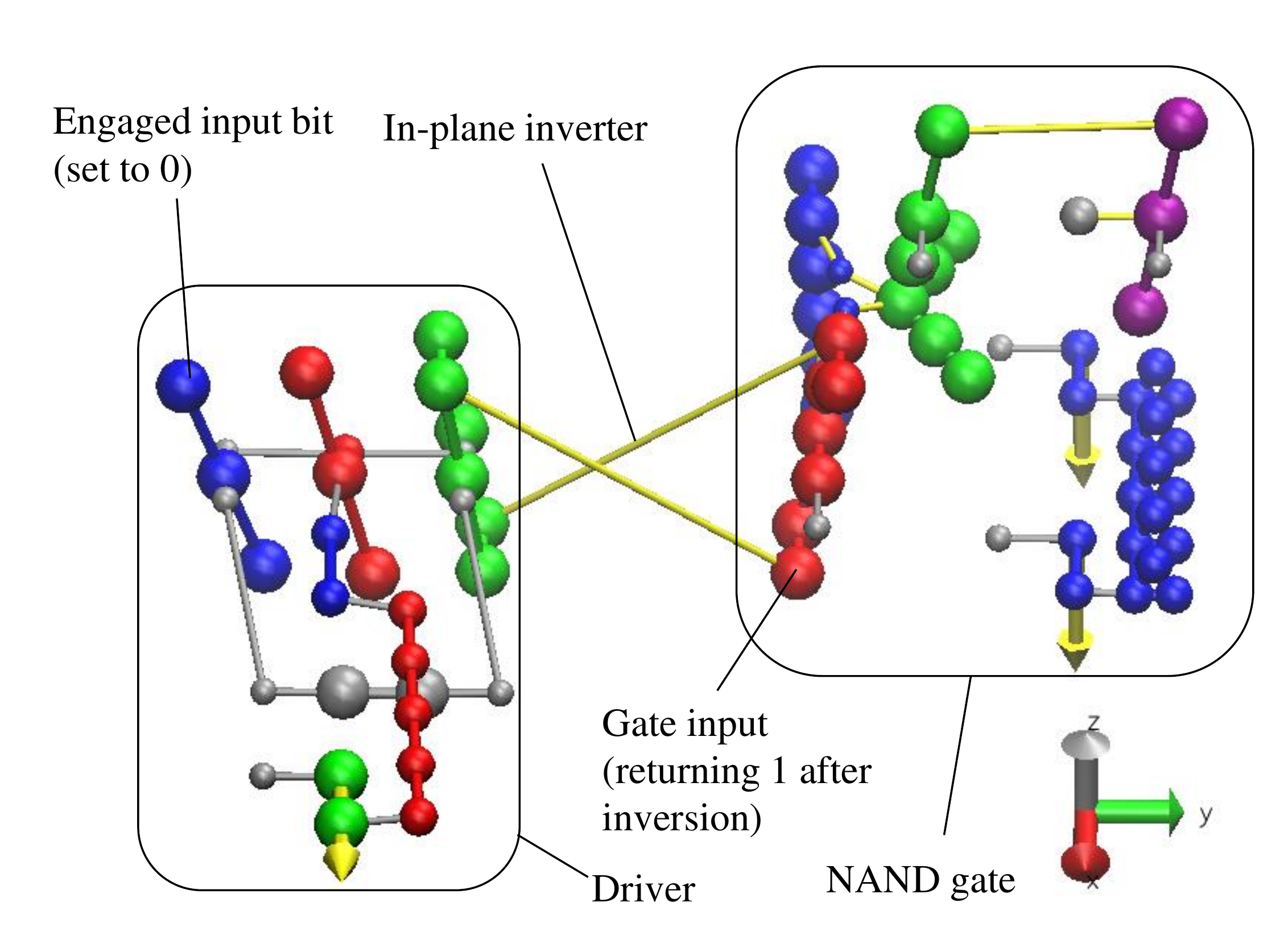}
	\caption{The output of a driver fed into the input of a NAND gate via an in-plane inverter.}
	\label{fig:ipi}
\end{figure}

The simple inverting connection does suffer from the disadvantage that the downstream gate or bit is not in the same plane, being displaced in the $z$-axis. When retaining both gates in the same plane is strongly desired for engineering reasons; for instance, in the half-adder circuit (Section \ref{ssec:hadder}), an alternative design where both upstream and downstream elements remain in the same plane can be used (Fig.\ \ref{fig:ipi}). By tethering the top of the upstream bit to the bottom of the downstream bit, increasing the slope of the input bit decreases the slope of the output bit via a pulling motion. However, decreasing the slope in a similar manner results in a pushing motion, which is less efficient at force transmission. Mitigating this problem is possible if a second tether is attached from the bottom of the output bit to the top of the input bit, converting deflections in both directions into pulling motions.

Although direct tethering is the simplest way of constructing combinatorial circuits, it also constrains the relative geometry of the two gates, with only a single degree of freedom along the $y$-axis. A free geometric relationship between the two logic gates in the $xy$-plane is relatively simple to implement in principle by extending the rigid body which  receives the input along the $x$-axis. 

\begin{figure}
	\includegraphics[width=\linewidth]{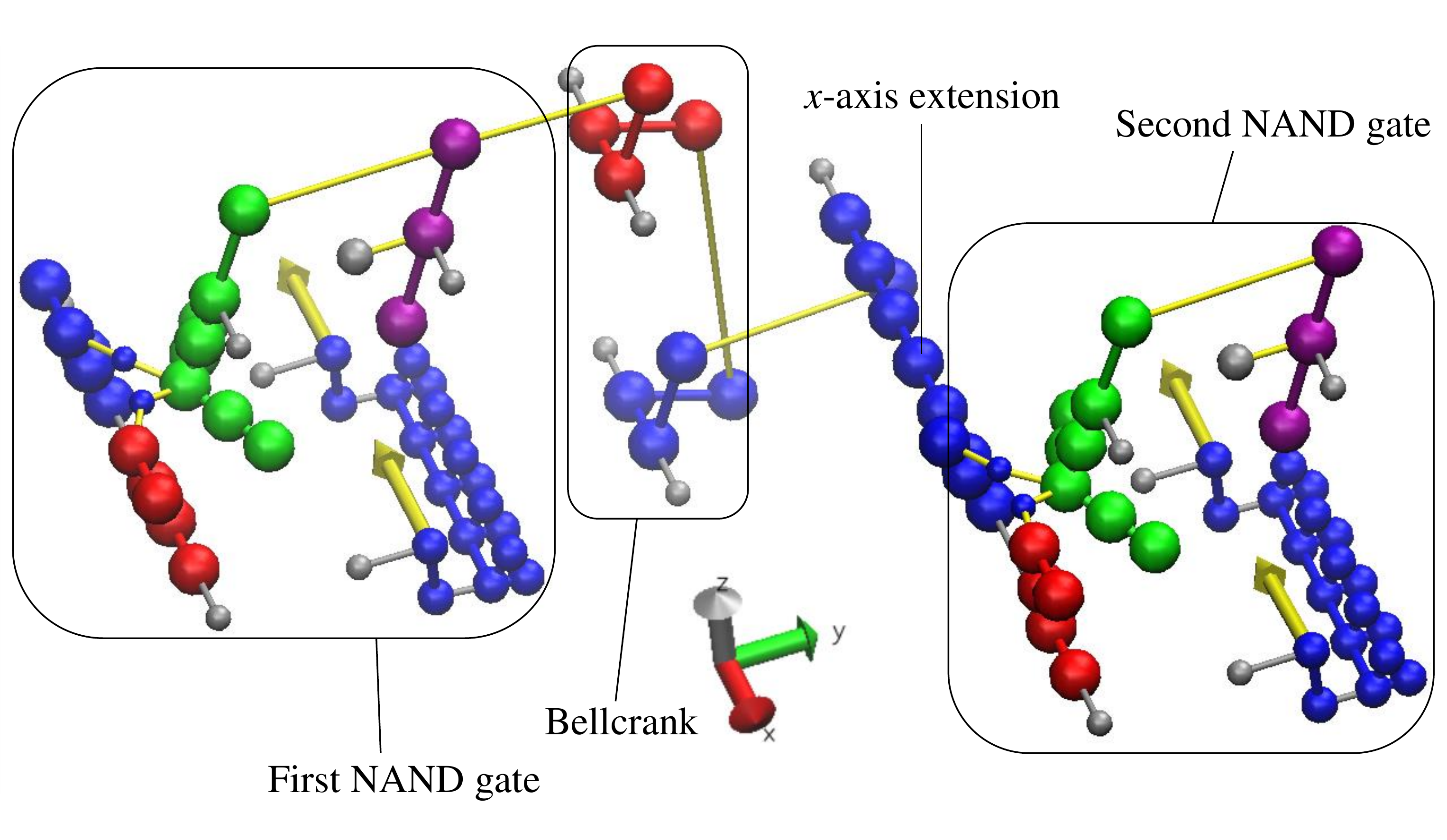}
	\caption{The NAND-inverter-NAND circuit from Fig.\ \ref{fig:inverter} further extended in the $y$ and $z$ directions  by the use of a bellcrank.}
	\label{fig:bellcrank}
\end{figure}

A free geometric relationship in all three axes is somewhat more complicated to achieve, as it requires the use of a bellcrank (Fig.\ \ref{fig:bellcrank}), a mechanism for converting a displacement along one axis into displacement along another orthogonal axis. The ability to construct three-dimensional circuits removes the fundamental constraint that the circuit graph be planar which afflicts two-dimensional circuits such as most CMOS-based designs. It should be noted that three-dimensional CMOS-based integrated circuits do exist; however, their practical use is generally limited to low-power applications due to heat dissipation issues \cite{san14}. The far greater theoretical efficiency of molecular mechanical logic gates would allow them to circumvent this limitation, potentially favouring three-dimensional molecular-mechanical logic circuits over two-dimensional circuits.

\section{\label{sec:landauer}Landauer's Principle and Szilard Engine}
The aim of the first set of simulations was to demonstrate Landauer's principle using the thermostat and driver. Other than demonstrating a fundamental physical principle, this set of simulations also serves to prove that the thermostat functions in a reasonable manner under the given conditions, and that persistent storage of information in a molecular mechanical system is possible.

\begin{figure}
	\includegraphics[width=\linewidth]{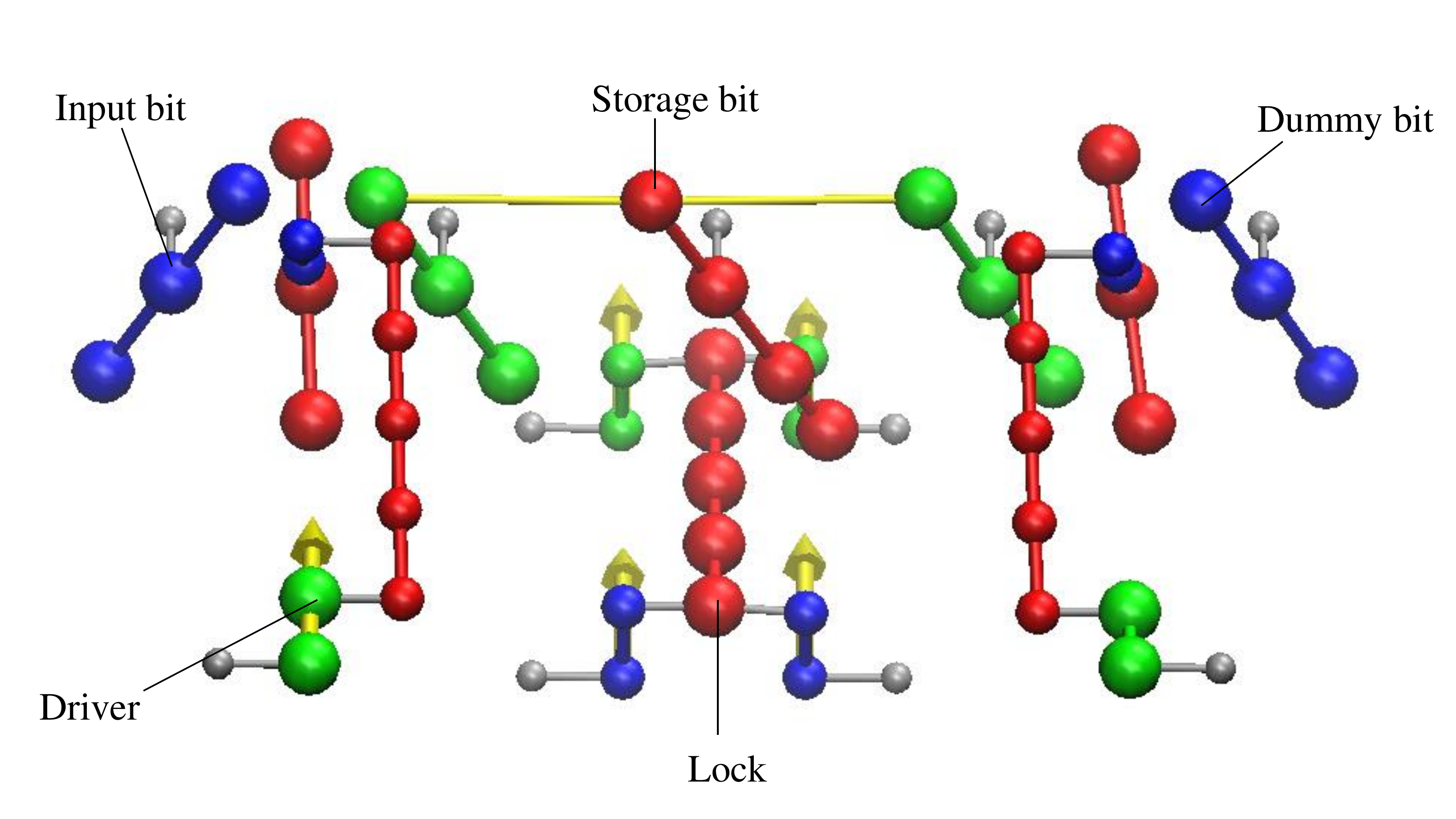}
	\caption{The driver-bit system used to demonstrate the Landauer bound and molecular Szilard engine with input bit initialised to `1' and storage bit to `0' at $t^*$ = 0 where $t^*$ is the current time fraction of the total time required to complete the full process.}
	\label{fig:doubleBit}
\end{figure}

In order to demonstrate Landauer's principle, an information storage system based on a simplified version of the driver outlined in Section \ref{ssec:driver} was designed. It consists of two drivers coupled to a single storage bit (Fig.\ \ref{fig:doubleBit}), though only one of the drivers responds to the movement of the external dipole. The driver design used is less complex than the one discussed in Section \ref{ssec:driver}, lacking the scaffold to force the driver arms to travel in only one direction; this was possible because there was no significant deviation from the most favourable path for the drivers of this particular system (a phenomenon that did not apply to the other systems simulated).

The storage bit receives information from the input bit via the left-handed driver only. The presence of the seemingly redundant right-handed driver ensures that the initial and final states in the case where the storage bit is overwritten are mirror images of each other. Therefore, the energy of the initial and final states are identical, a fact which greatly simplifies the task of demonstrating the Landauer limit.

Another significant component of the double-bit system is the lock. This is a rod which when engaged acts to restrict the motion of the storage bit, forcing it to remain in one of two possible states until the lock is disengaged. The motion of the lock is driven by a pair of synchronised internal dipoles. Despite superficially resembling the signal booster used to augment the signal of the winged NAND gate, the internal dipoles which drive the lock do not rotate synchronously with the internal dipole of the left-handed driver. Therefore, lock engagement and disengagement is not a logically reversible operation. This logical irreversibility is a direct consequence of the fact that during the reverse process, the lock becomes engaged before the driver does, causing it to force the bit into a random state.

\begin{figure}
	\includegraphics[width=\linewidth]{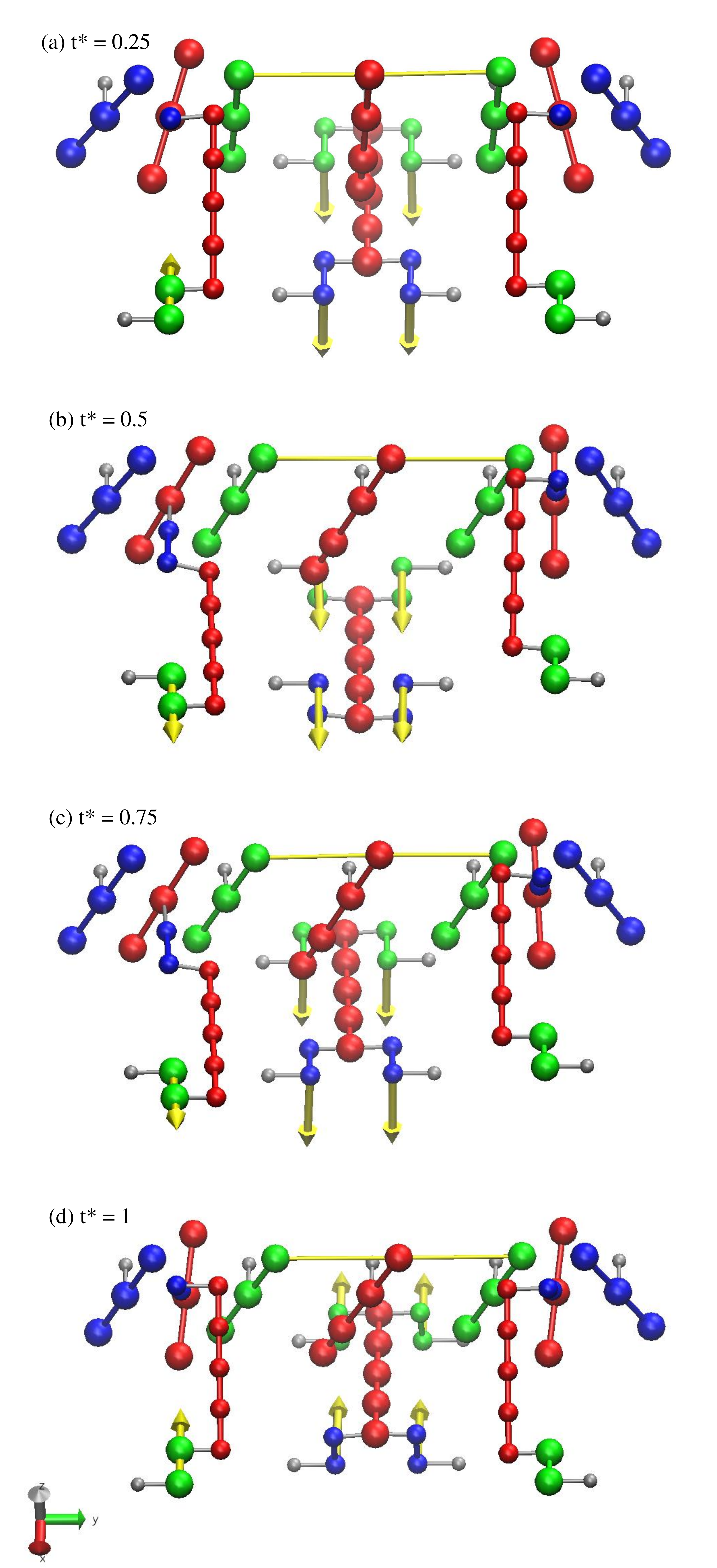}
	\caption{The driver-bit system between $t^* = 0.25$ and $t^* = 1$ during an overwriting process. {Supplemental Video 2 provides an animation of this process.}}
	\label{fig:doubleBitGallery}
\end{figure}

Fig.\ \ref{fig:doubleBitGallery} demonstrates the process of writing information from the input bit to the storage bit from time $t^* = 0.25$ where $t^*$ is the ratio of the current time $t$ over the time required to complete one full clock cycle $\tau$. The system is initialised with lock engaged and driver disengaged, with the internal dipole of both driver and lock $\hat{\mathbf{d}}_\mathrm{lock} = \hat{\mathbf{d}}_\mathrm{driver} = (-1, 0, 0)$. As the system moves from $t^* = 0$ to $t^* = 0.25$, $\hat{\mathbf{d}}_\mathrm{lock}$ is rotated by $90\degree$ to $(0, 0, -1)$, causing the lock to become partially disengaged. Between $t^* = 0.25$ to $t^* = 0.5$ the internal dipole of the driver rotates by $180\degree$ to $(1, 0, 0)$, causing information to be transduced from the input bit to the storage bit. Simultaneously, the internal dipole of the lock rotates by $90\degree$ to $(1, 0, 0)$, though this has no effect on the storage bit since it is already free to move. Between $t^* = 0.5$ to $t^* = 0.75$ the internal dipole of the lock is rotated back to $(0, 0, -1)$, though as before this has no effect on the storage bit. Finally, between $t^* = 0.75$ to $t^* = 1$ the lock is engaged as its internal dipole is rotated back to its initial position, preventing the storage bit from moving freely. Simultaneously, the internal dipole of the driver is rotated back to its initial position, disengaging the input and storage bits.

Note that the dipoles of the driver and the lock are driven by separate external dipoles in this system, unlike the signal booster of the NAND gate, hence the reason they are out of phase. This process causes data in the storage bit to be overwritten, and is thus logically irreversible due to reducing the information entropy of the system. Therefore, it is expected that the process will require $k_\mathrm{B}T\ln2$ of work to complete in the quasistatic limit.

In addition to demonstrating the work required to overwrite a bit, the driver-bit system can also be used to extract work from a system whose state is known. This process resembles the second phase of the set-up of a well-known thought experiment, the Szilard engine \cite{szilard}, in which usable work is extracted from information. Similarly, by reversing the process in which information is written into the storage bit, usable work may be extracted from the system at the cost of randomising the state of the storage bit due to the thermodynamically reversible transfer of the potential energy contained in the high-information (low-entropy) state into the external dipoles.  Thus, the full process of overwriting a bit of randomised state by expending work followed by the recovery of the work done at the cost of overwriting the bit is an example of a logically irreversible process being executed in a thermodynamically reversible manner. 

\begin{figure}
	\includegraphics[width=1\linewidth]{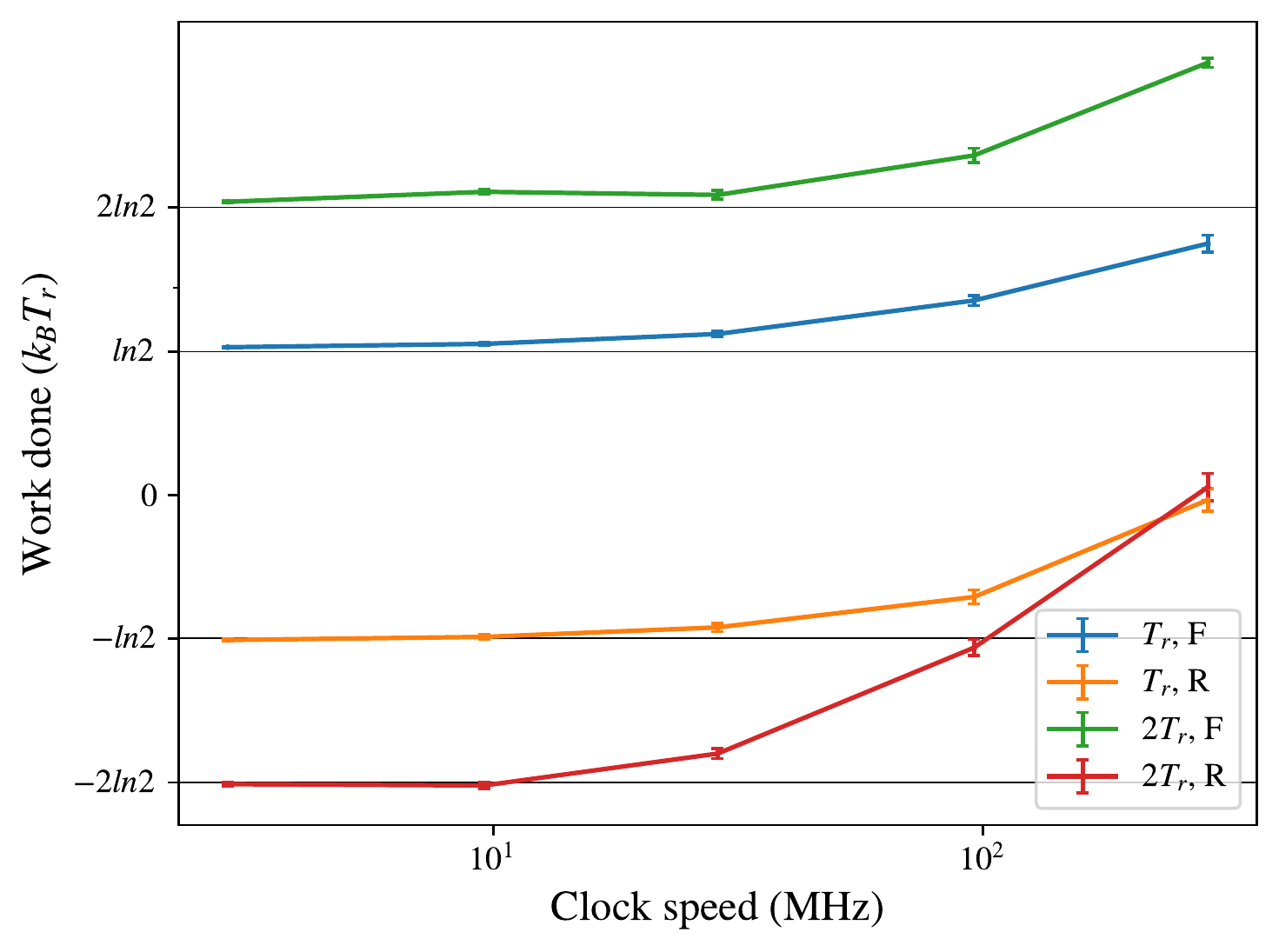}
	\caption{Work done against clock speed at two different temperatures for the forward and reverse bit-writing processes of the double-bit system. The blue and orange lines correspond to the forward (F) and reverse (R) simulations carried out at $T_{r}$ (298\,K), respectively, while the green and red lines correspond to those carried out at $2T_{r}$ (596\,K). Each data point corresponds to an average over 240 simulations.}
	\label{fig:dsin}
\end{figure}

Two sets of simulations were carried out, corresponding to the situation where the input bit is initialised to `0' and `1'. The work done for both sets of simulations was averaged together and the results plotted in Fig.\ \ref{fig:dsin}. From the graph, it is clear that reducing the clock speed causes the work done to converge to $k_\mathrm{B}T\ln2$ at both room temperature ($T_{r}$ = 298\,K) and 596\,K. In addition, the work recovered by the reverse process also converges to $k_\mathrm{B}T\ln2$ as the clock speed is reduced, indicating that the system is capable of recovering the free energy contained in high-information states.

\section{\label{sec:nand}Winged NAND gate}

\subsection{Neutral transitions}
In order to determine the viability of the winged NAND gate design, reversibility and error rate simulations were run using the setup outlined in Fig.\ \ref{fig:NANDgallery}. In these simulations, the system was initialised with both drivers at a point of minimum engagement. Before each simulation was run, the system was driven over one full cycle and then re-equilibrated to ensure that the starting state was as close to thermodynamic equilibrium as possible. The drivers were then rotated by the external dipole to the point of maximum engagement at $t^* = 0.5$. The system was then returned to the initial state by a reverse rotation of the external dipole at the same rate; the full process constitutes one clock cycle.  Animations of the NAND gate in the `00', `01' and `11' states driven via the process described above can be viewed in Supplemental Videos 3-5.

Due to the lack of potential energy barriers in the disengaged (neutral) state, it can be assumed that the free energy of the initial and final states are practically equal. The error rate was calculated by taking the ratio of simulations where the output bit returned an incorrect value at the point of maximal driver engagement to the total number of simulations carried out. $120\,000$ error-rate simulations were run at 1.44\,GHz and at temperatures of 1, 2, 4 and $6T_{r}$; of these, only the simulations run at 6\,$T_{r}$ had non-zero error rates. The results of these simulations are summarised in Table \ref{tab:nandNeut}. The error rate was below $8.33 \times 10^{-6}$ at 1, 2 and $4T_{r}$, and still relatively low at $6T_{r}$. Given the relatively low error rate even at highly elevated temperatures, it is almost certain that the error rate at 298\,K would be sufficiently low so as not to interfere with demonstrations of thermodynamic reversibility.

\begin{figure}
	\includegraphics[width=1\linewidth]{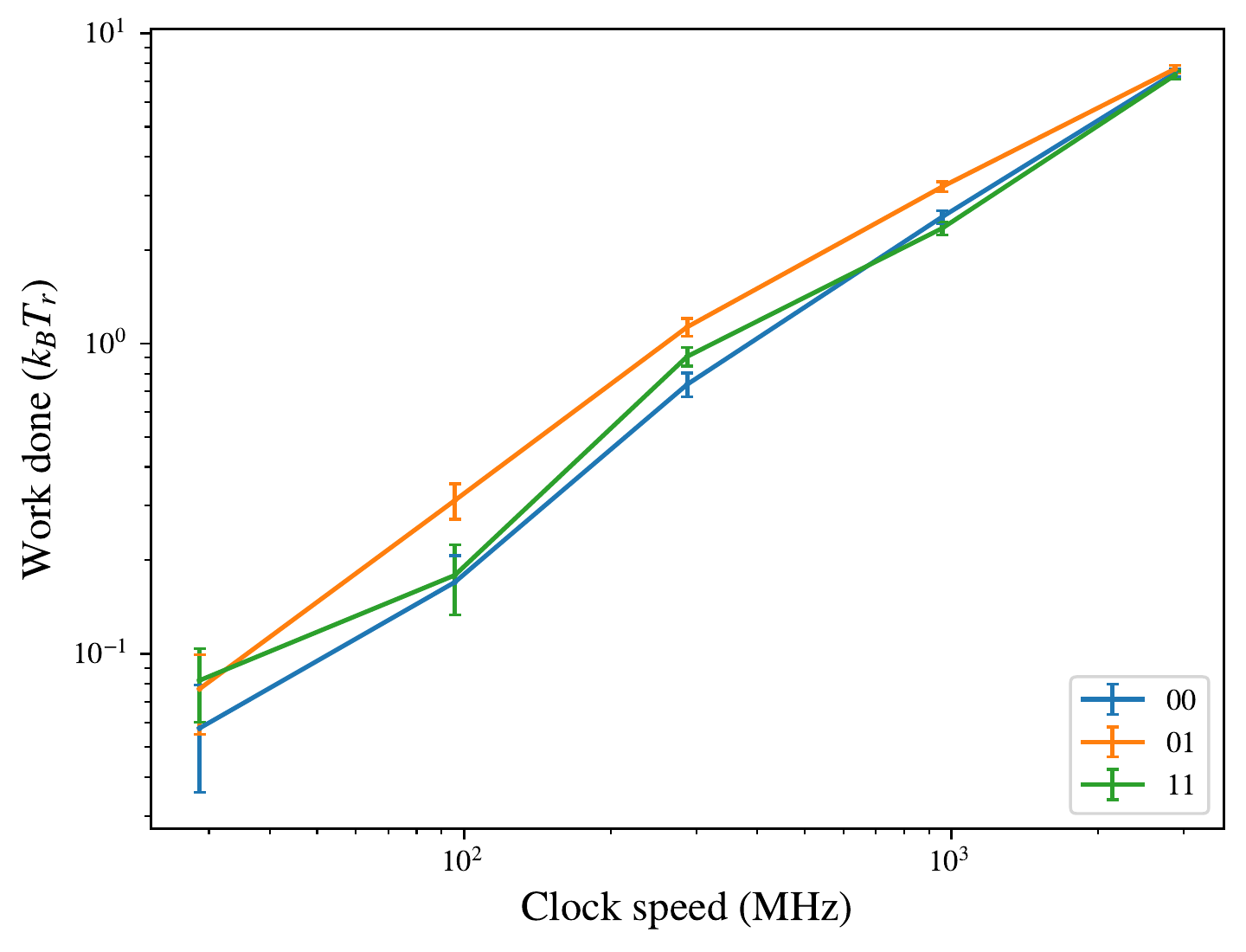}
	\caption{Work done against clock speed for the NAND gate during thermodynamic reversibility simulations from a neutral state to the state to which the input bits are set. Each data point corresponds to an average over 120 simulations.}
	\label{fig:neut}
\end{figure}

A second set of simulations was run to determine the thermodynamic reversibility of the design. The results are summarised in Fig.\ \ref{fig:neut}. As the work done by the external dipole is proportional to  clock speed  for all states, it can be concluded that this system converges toward thermodynamic reversibility in the limit of clock speeds tested.

\begin{table}
\caption{\label{tab:nandNeut}Results obtained from error-rate simulations of the NAND gate performed at $6T_{r}$ (1788\,K) for the evolution from a neutral state to a given target state at a clock speed of 1.44\,GHz. No errors were detected over $120\,000$ simulations at $T \leq 4T_{r}$.}
\begin{tabular}{c@{\hspace{1cm}}c}
\midrule
\midrule
State & Error rate\\
\midrule
00 & 3.33$\times10^{-5}$\\
01 & 2.17$\times10^{-4}$\\
10 & 5.83$\times10^{-5}$\\
11 & 5.00$\times10^{-5}$\\
\midrule
\midrule
\end{tabular}
\end{table}

\subsection{Switching between defined states}
\begin{figure}
	\includegraphics[width=\linewidth]{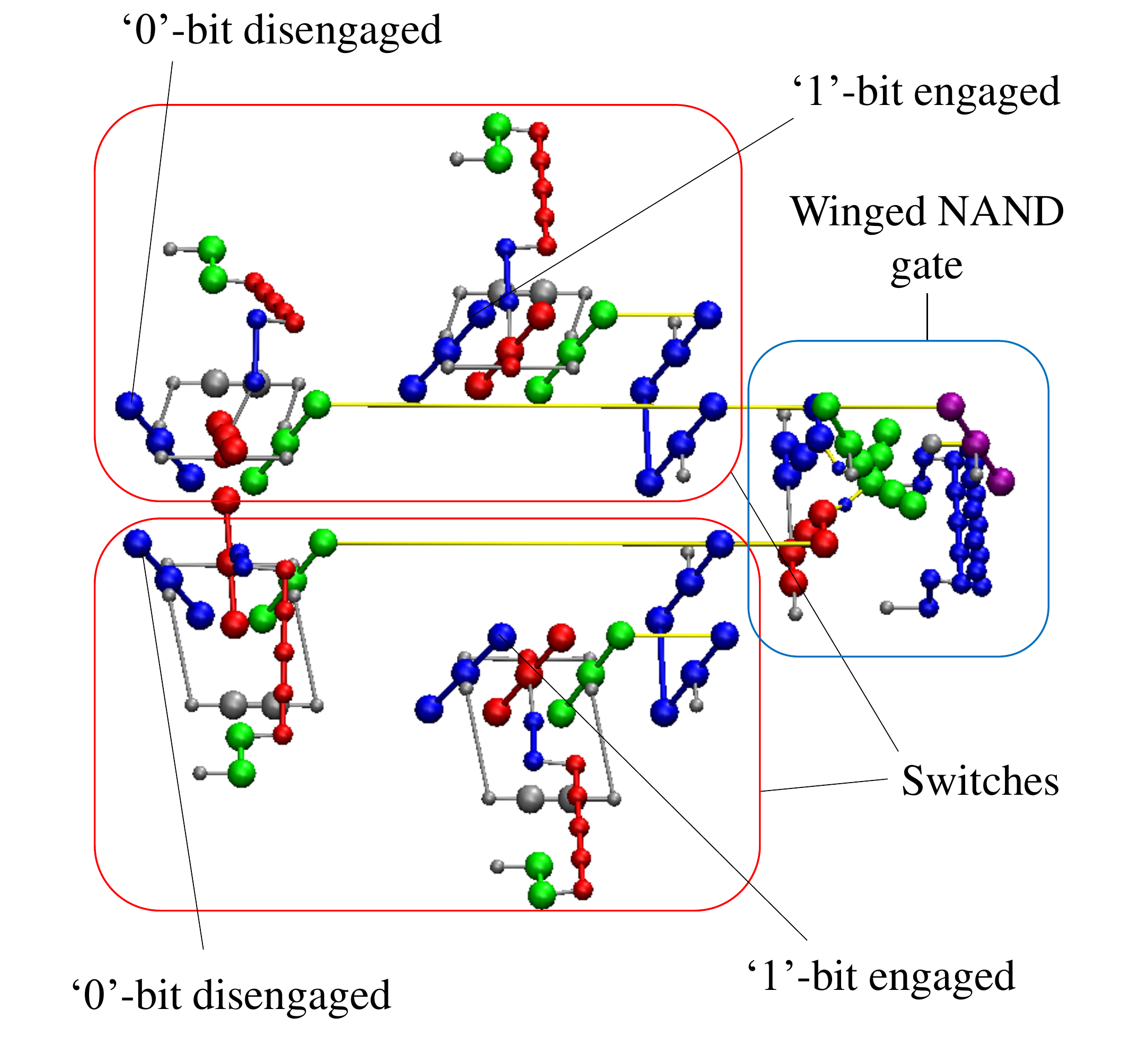}
	\caption{Two pairs of switches (red outline) connected to a winged NAND gate (blue outline).}
	\label{fig:nandSwitch}
\end{figure}
Switching between defined states (states in which the bits are constrained in place due to driver engagement) and monitoring the error rate of both the initial and final states of the system after a round-trip process provides an even stronger guarantee that the initial and final states are identical in free energy, so long as the error rate remains extremely low at the desired operating temperature. In order to carry out these simulations, the set-up in Fig.\ \ref{fig:nandSwitch}, consisting of two pairs of switches coupled to a NAND gate, was used to drive the NAND gate between all twelve possible pairs of inputs and outputs.

\begin{figure}
	\includegraphics[width=1\linewidth]{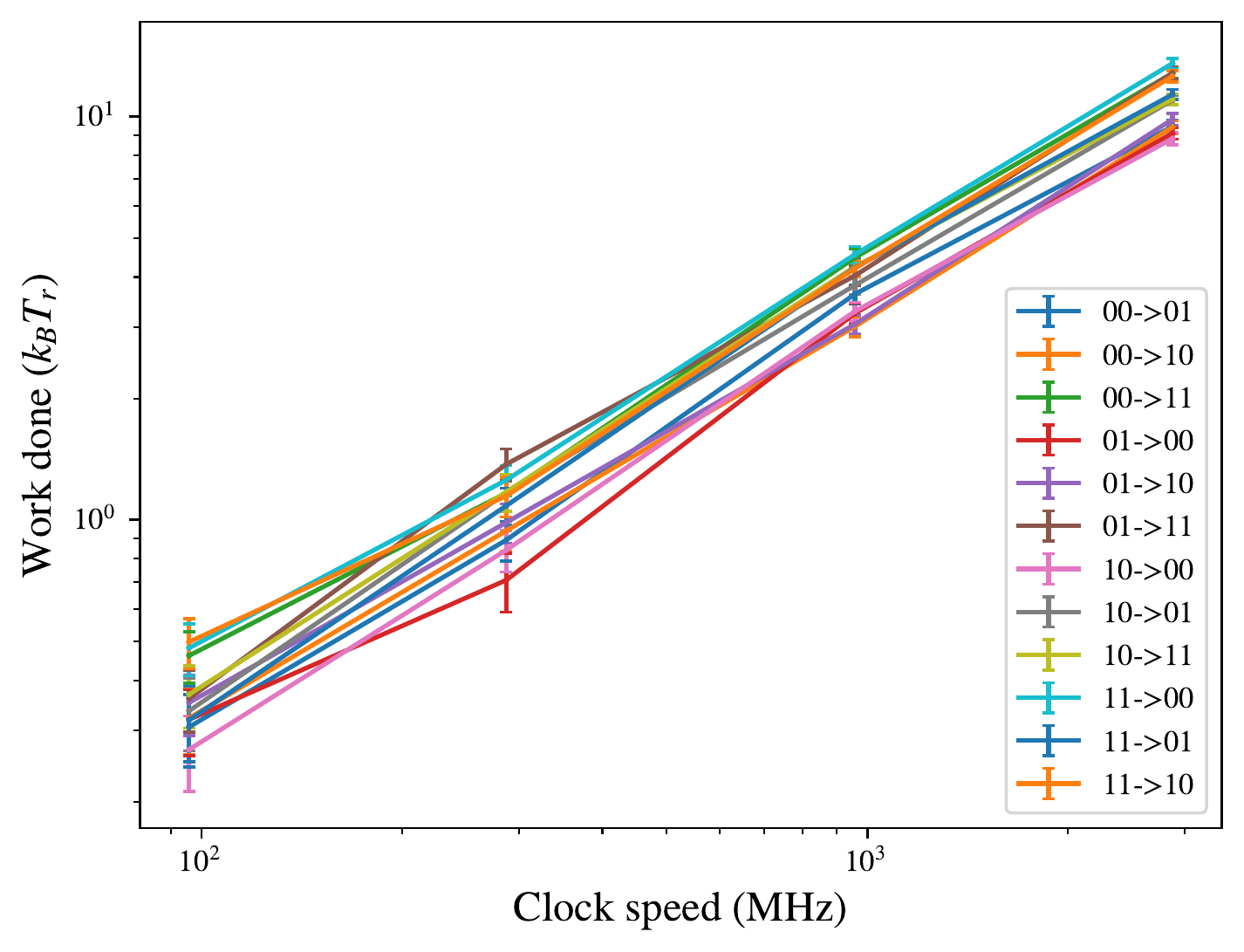}
	\caption{Work done against clock speed for the NAND gate during thermodynamic reversibility simulations of transitions from one defined state to another. Each data point corresponds to an average over 120 simulations.}
	\label{fig:mux}
\end{figure}

In this protocol, the system is initialised with the drivers of the switch engaged towards the initial state and the signal booster of the NAND gate fully engaged. The starting error in Table \ref{tab:muxNand} corresponds to the error rate of this initial state (measured after equilibration); thus, if it is assumed that the system is transitioning from `11' to `00' as in Fig.\ \ref{fig:mux}, this corresponds to the error rate of the NAND gate at `11'. The drivers of the switch rotate in the same direction, but at half the rate of the signal booster of the NAND gate.

Between $t^* = 0$ and $t^* = 0.25$, the signal booster of the NAND gate is rotated by $180\degree$ to become fully disengaged while the switch drivers rotate by $90\degree$, leaving the input bits in an indeterminate state. Between $t^* = 0.25$ and $t^* = 0.5$, the signal booster is rotated in the same direction by another $180\degree$, becoming re-engaged; simultaneously, the switch drivers rotate by another $90\degree$ causing the drivers of the switch to engage the second set of bits. It is at this point that the midpoint error is measured; this corresponds to the error rate of the `00' state in Fig.\ \ref{fig:mux}. This process is then reversed, with the endpoint error corresponding to the error rate of the `11' state for this system after it has been returned to its initial configuration at $t^* = 1$. 
 Supplemental Video 6 provides an animation of such a process.
 As with the neutral transition simulations, the energy-minimised system was driven over one full cycle and then re-equilibrated before any observables were measured.

The results of reversibility and error rate simulations, recorded in Fig.\ \ref{fig:mux} and Table\ \ref{tab:muxNand} respectively, show that the system tends towards thermodynamic reversibility across all possible unique transitions in the limit of clock speeds tested and has a reasonably low error rate even at $4T_{r}$, sufficiently low that it is fair to assume that the initial states of all transitions at 298\,K are in equilibrium.

\begin{table}
\caption{\label{tab:muxNand}Results obtained from error-rate simulations of the NAND gate performed at $4T_{r}$ (1192\,K) for the evolution from a defined starting state to a given target state at a clock speed of 1.44\,GHz. No errors were detected over $60\,000$ simulations at $T \leq 3T_{r}$.}
\begin{ruledtabular}
\begin{tabular}{cccc}
Transition & Error (Start) & Error (Mid) & Error (End) \\
\midrule
\centering
00$\rightarrow${}01 & 1.667$\times10^{-5}$ & 5.000$\times10^{-5}$ & 1.667$\times10^{-5}$ \\
00$\rightarrow${}10 & 0 & 3.333$\times10^{-5}$ & 0 \\
00$\rightarrow${}11 & 0 & 0 & 0 \\
01$\rightarrow${}00 & 0 & 0 & 0 \\
01$\rightarrow${}10 & 3.333$\times10^{-5}$ & 5.000$\times10^{-5}$ & 6.667$\times10^{-5}$ \\
01$\rightarrow${}11 & 1.500$\times10^{-4}$ & 3.333$\times10^{-4}$ & 1.167$\times10^{-4}$ \\
10$\rightarrow${}00 & 0 & 0 & 3.33$\times10^{-5}$ \\
10$\rightarrow${}01 & 6.667$\times10^{-5}$ & 6.667$\times10^{-5}$ & 6.667$\times10^{-5}$ \\
10$\rightarrow${}11 & 1.000$\times10^{-4}$ & 0 & 1.000$\times10^{-4}$ \\
11$\rightarrow${}00 & 0 & 0 & 0 \\
11$\rightarrow${}01 & 3.167$\times10^{-4}$ & 5.000$\times10^{-4}$ & 3.333$\times10^{-4}$ \\
11$\rightarrow${}10 & 0 & 1.000$\times10^{-4}$ & 0
\end{tabular}
\end{ruledtabular}
\end{table}

\section{\label{sec:chain}Chaining Gates}

\subsection{\label{ssec:dirChain}Direct chaining}
As previously mentioned, the simplest way to chain two gates together is to directly feed the output of one into the input of another, a process henceforth referred to as direct chaining. Despite its simplicity, directly chaining input-preserving molecular mechanical gates together to create logical circuits is not always practical due to the high error rate which may result.
\begin{figure}
	\includegraphics[width=\linewidth]{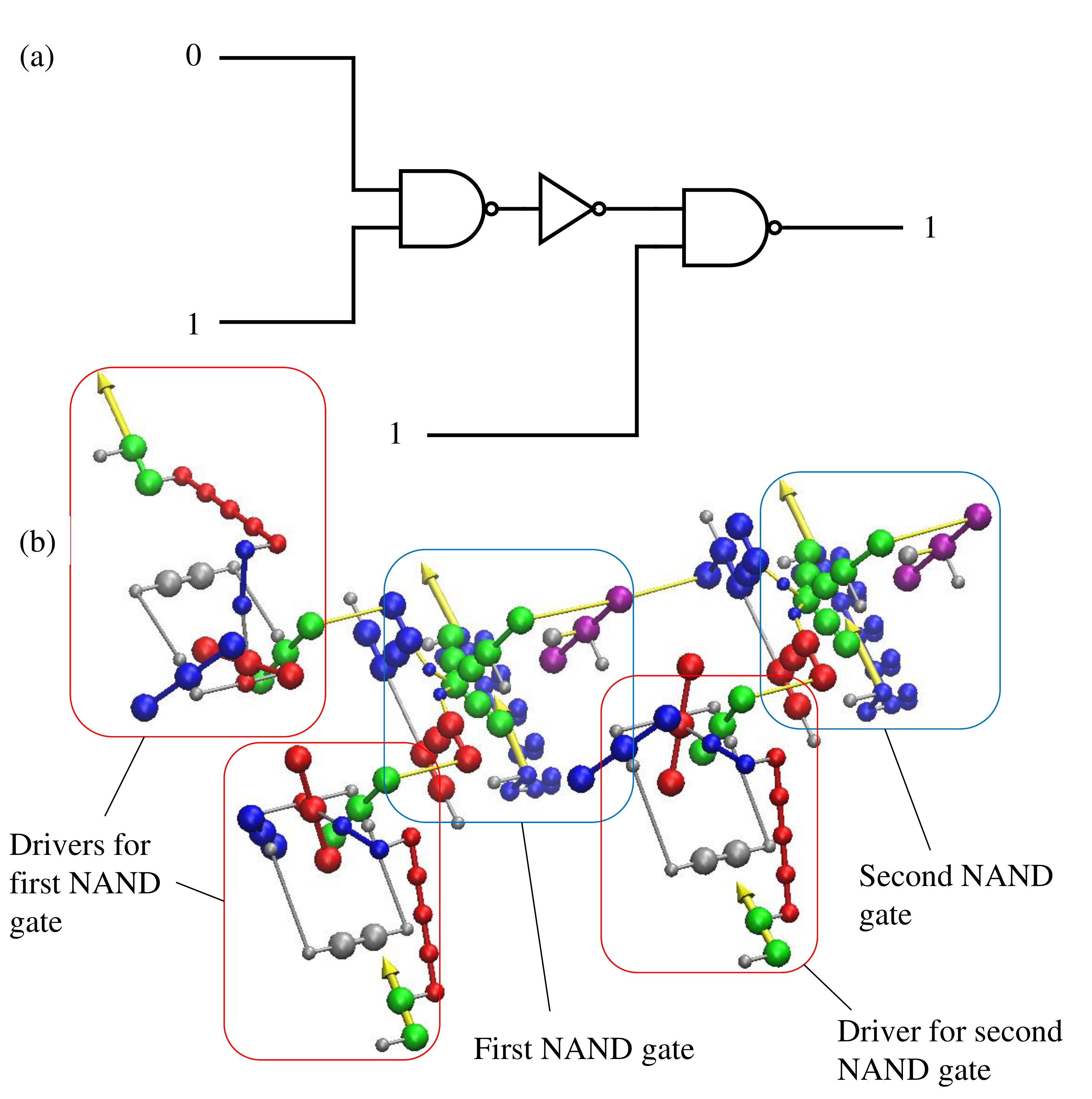}
	\caption{The circuit layout used to demonstrate direct gate chaining.}
	\label{fig:directChainGallery}
\end{figure}

In order to demonstrate the error rate of direct chaining, the circuit shown in Fig.\ \ref{fig:directChainGallery} was used. This particular circuit can be easily extended by the addition of more inverter-NAND units since each NAND gate receives an input of `01' and outputs `1', which is inverted and passed along to the next NAND gate in line.

Only one set of error-rate simulations were run for the direct-chaining system. In these simulations, the system was initialised with all drivers and signal boosters fully disengaged. As all drivers and signal boosters are by definition in phase, they simultaneously become fully engaged at $t^* = 0.5$, at which point the error rate of the final gate in the chain is measured. The process is then reversed.

Error rates of 4.6\% and 12.9\% were obtained for directly-chained systems of two and three gates, respectively. The error rates, measured at room temperature, are far higher than that for a single NAND gate, and demonstrate the predicted problems of direct chaining. No thermodynamic reversibility simulations were attempted due to the high error rate.

We propose two major causes for the large increase in error rate as a result of directly chaining two gates together. Firstly, the spring constant of a number of identical simple harmonic oscillators in series is inversely proportional to the number of oscillators linked together. Similarly, a long series of chained gates becomes increasingly less able to effectively transmit information from the first gate to the last without errors caused by thermal fluctuations. Secondly, due to the finite rigidity of the system, there is insufficient time for the information from the first gate in the chain to reach subsequent gates in the chain before the associated signal boosters is activated (the activation occurring simultaneously with that of the first gate), causing the signal booster to occasionally force the output of the second gate into an erroneous position.

\subsection{\label{ssec:phaseChain}Phased chaining}
\begin{figure}
	\includegraphics[width=\linewidth]{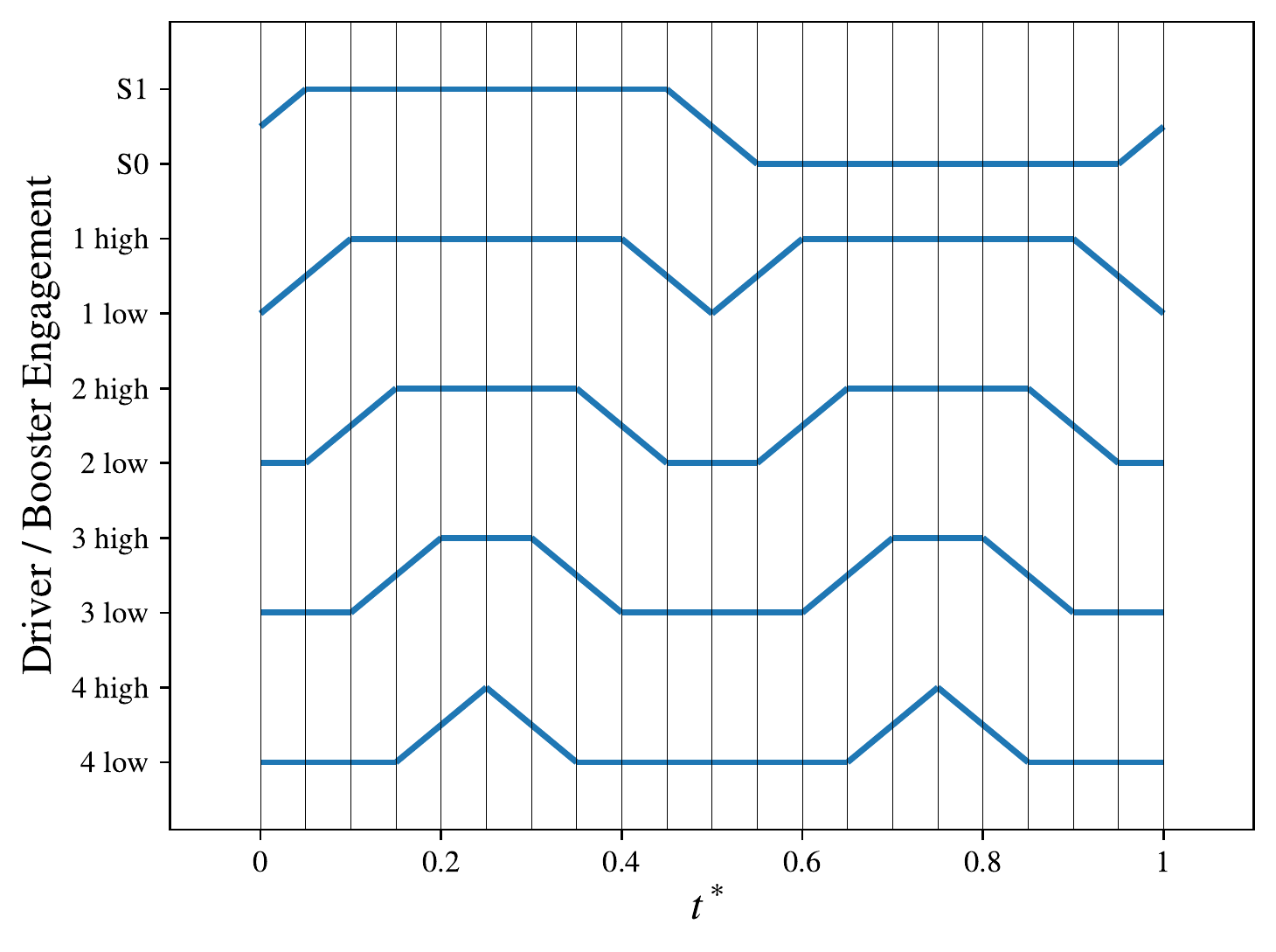}
	\caption{Illustration of the phased chaining protocol for a system containing four gates. The $n$-high/low labels on the $y$-axis indicate whether the drivers or signal boosters of the $n$th gate are engaged. S0 and S1 indicate the state of the switches when fully toggled toward `0' and `1' respectively. Each gate is engaged in turn and only disengaged after all succeeding gates have already been disengaged.}
	\label{fig:phaseGraph}
\end{figure}

A more sophisticated approach is required to ensure acceptable error rates and reversibility when chaining input-preserving gates together. One method of reducing error rate is to boost the signal at each gate while staggering the boosts so they are out of phase as shown in Fig.\ \ref{fig:phaseGraph}. In addition, a driver is placed between the output of an upstream gate and the input of the downstream one, allowing the two to be physically separated until the upstream gate's signal has been fully boosted. This protocol should, in principle, allow each gate to be treated as an independent unit.

This protocol does require individual control over each distinct level of logic gates, making it impossible to control with a single external dipole. In addition, the rotation associated with each level must be halted at the point of maximal engagement; this precludes the use of a continuously rotating dipole. This limitation is not necessarily less realistic; extant molecular motors, both biological (e.g. the mitochondrial proton pump \cite{fried21}) and artificial \cite{hern04} generally rotate in discrete steps as opposed to supplying a continuous torque.

\begin{figure}
	\includegraphics[width=\linewidth]{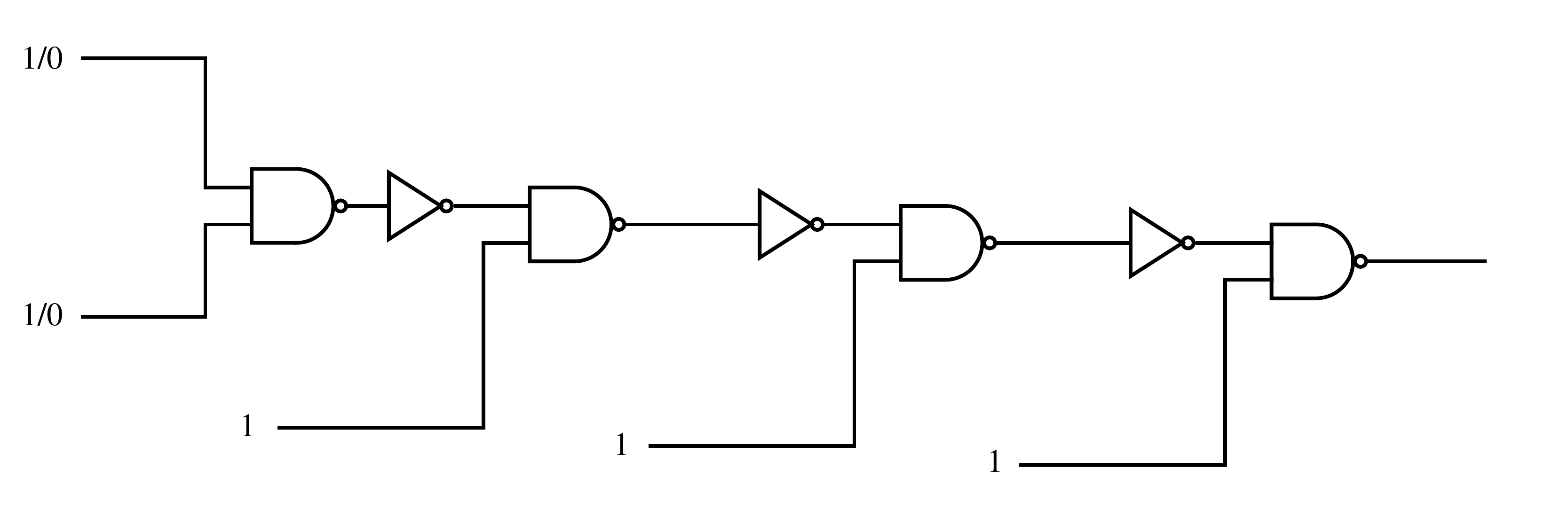}
	\caption{A circuit used to demonstrate phased chaining. This particular circuit has a chain length of 4 and can be extended indefinitely by the addition of more inverter-NAND units with the output of each NAND gate switching from `0' to `1' as the inputs of the starting NAND gate are switched from `11' to `00'.}
	\label{fig:phaseChainCircuit}
\end{figure}

The circuit layout in Fig.\ \ref{fig:phaseChainCircuit} was implemented to demonstrate this protocol, as it, like the circuit in Fig.\ \ref{fig:directChainGallery} has the advantage of allowing additional inverter-NAND units to be added to extend it indefinitely. The inputs of the first NAND gate in the chain is initialised to the state `11', and is shifted to the state `00' at the midpoint of the simulation by the switch; this process is then reversed to return the inputs to `11'.

To illustrate the phased chaining operation, we will use the simplest example with only two NAND gates. The protocol proceeds in two phases, the first in which the switches return an output of `1', and the second where the switches return an output of `0'. The system is initialised with the switch in the midpoint position (biased toward neither bit) and every gate and driver maximally disengaged at $t^* = 0$. As soon as the first phase is initiated, the dipoles of the switches and the signal booster of the first NAND gate begin to rotate, such that at $t^* = 1/12$, both switches now return `1'. At this point, the switch is halted. The dipole of the signal booster of the first NAND gate continues to rotate until it reaches maximal engagement at $t^* = 1/6$. Thus, the first NAND gate now outputs `0'. It is at this point that the first-phase error rate of gate 1 is measured. The driver and the signal booster of the second NAND gate begin and end their rotation in phase at $t^* = 1/12$ and $t^* = 1/4$ respectively, returning an output of `0'. It is at this point that the first-phase error rate of gate 2 is measured. Between $t^* = 1/4$ and $t^* = 1/2$, this process is reversed, returning the system to its initial state. The second phase is initiated at this point. It is identical to the first phase except that the switches are rotated in the opposite direction such that they return `0'. The second-phase error of gate 1 is thus measured at $t^* = 2/3$ and that of gate 2 at $t^* = 3/4$.

A similar protocol applies for chains containing three or more gates, with the $n$th NAND gate reaching maximal engagement in the first phase at $t^* = (2n - 1)/4(x + 1)$ where $x$ is the number of gates in the chain, and those in the second phase at $t^* = (2x + 2n + 1)/4(x + 1)$. It is at these points that the error rate is measured. A chain with $n$ gates requires $n + 1$ independent external dipoles to drive it. One clock cycle is defined as $t/2(x + 1)$, that is, twice the time needed for the signal booster of one gate to rotate from the disengaged to engaged position; this corresponds well with the clock cycle definitions of earlier simulations.  Animations of the phased-chained systems of lengths 2, 3 and 4 can be viewed in Supplemental Videos 7, 8 and 9, respectively.

\begin{figure}
	\includegraphics[width=1.05\linewidth]{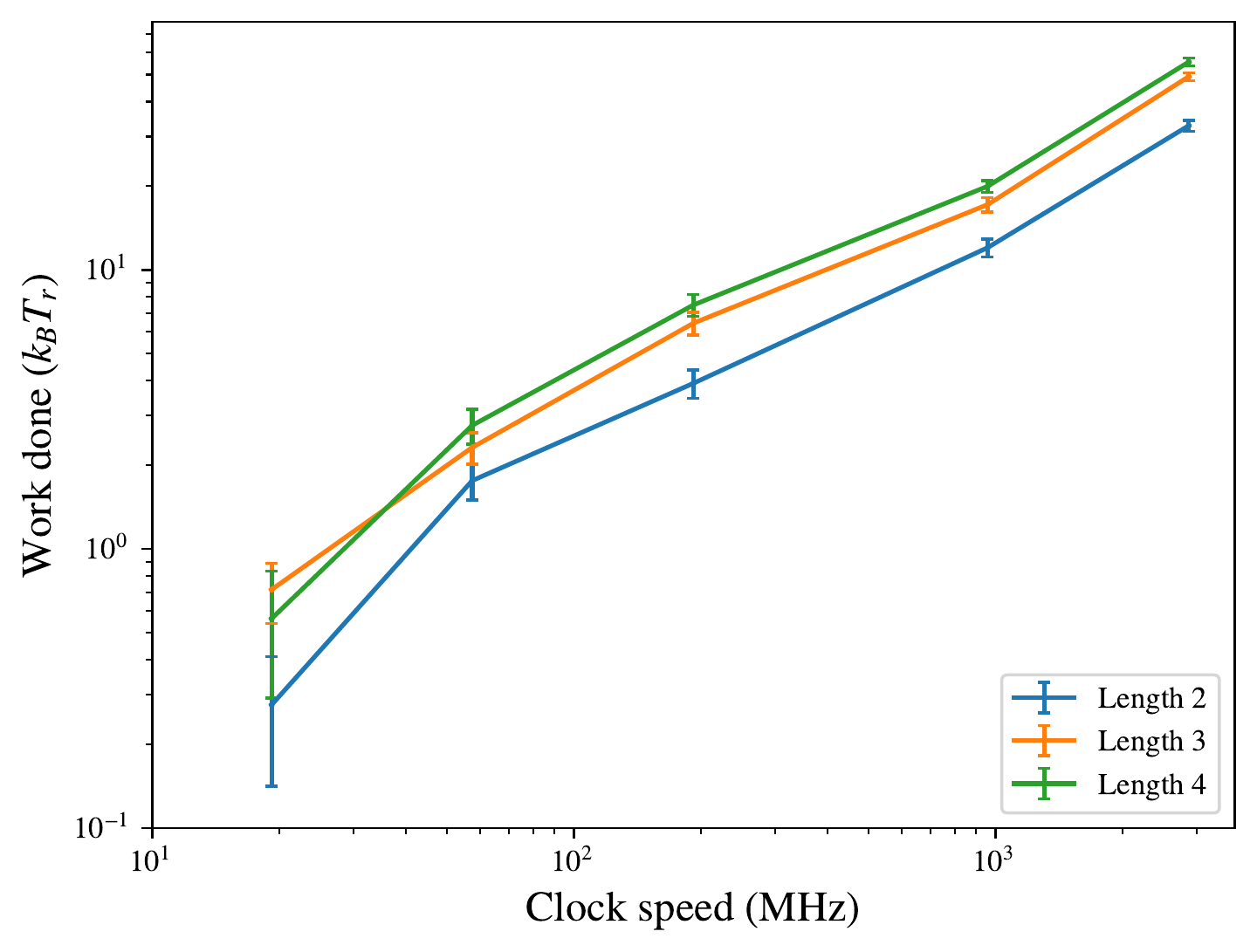}
	\caption{Plot of work done against clock speed for different chain lengths in phased chaining. Each data point corresponds to an average over 120 simulations.}
	\label{fig:muxRev}
\end{figure}

The results of the reversibility and error rate simulations are summarised in Fig.\ \ref{fig:muxRev} and Table \ref{tab:phaseChain}. The reversibility simulations show an approximately linear relationship between work done and clock speed, while the error rate remains low even at $6T_{r}$. Overall, it appears reasonable to conclude that this method of chaining input-preserving NAND gates is tending towards thermodynamic reversibility in the limit of clock speeds tested.

\begin{figure*}
    \includegraphics[width=\linewidth]{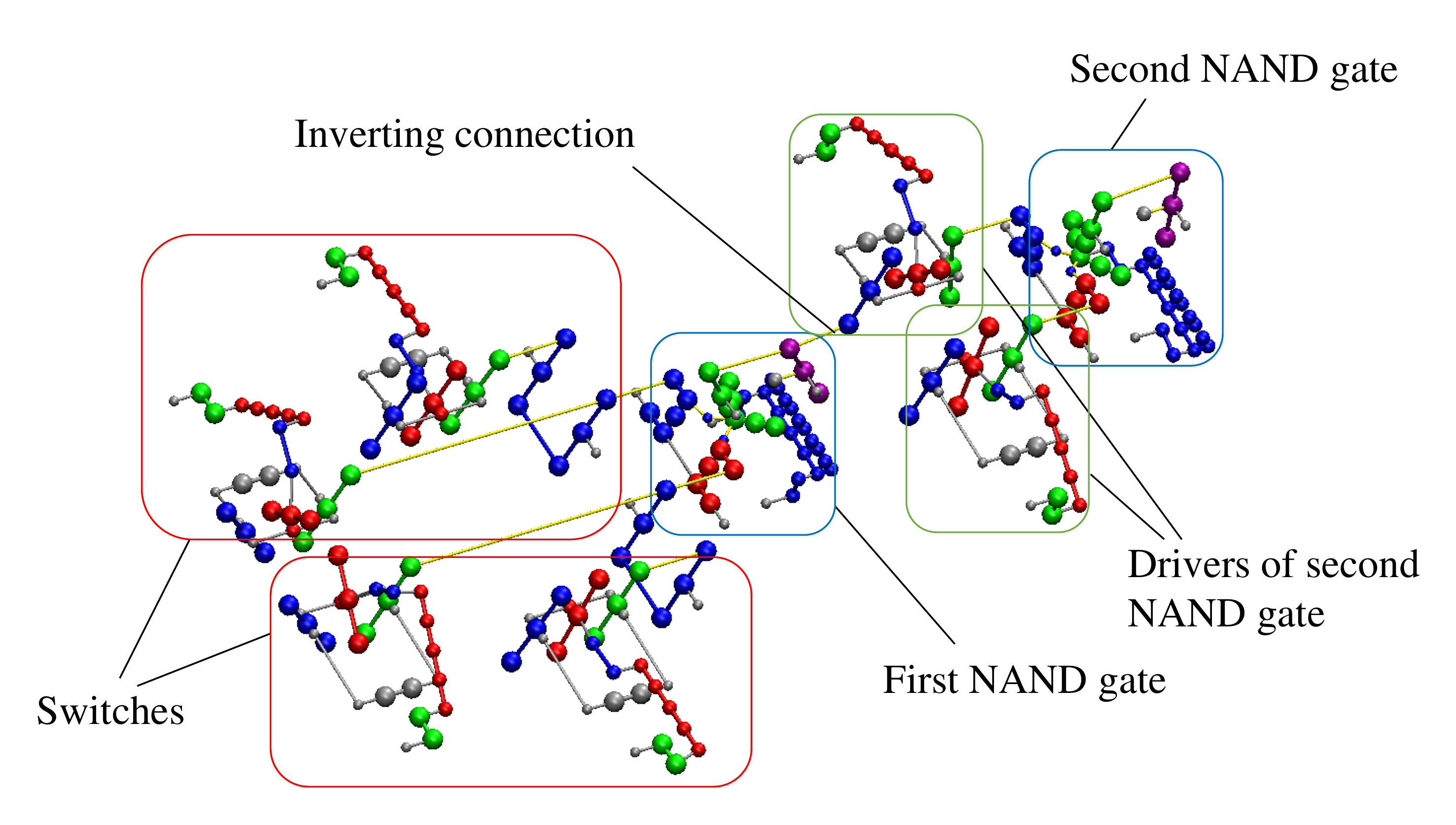}%
	\caption{ A phased-chained system consisting of one initial NAND gate and one inverter-NAND gate unit (chain length 2).}
	\label{fig:phaseChainLabel}
\end{figure*}
\begin{figure}
	\includegraphics[width=1.03\linewidth]{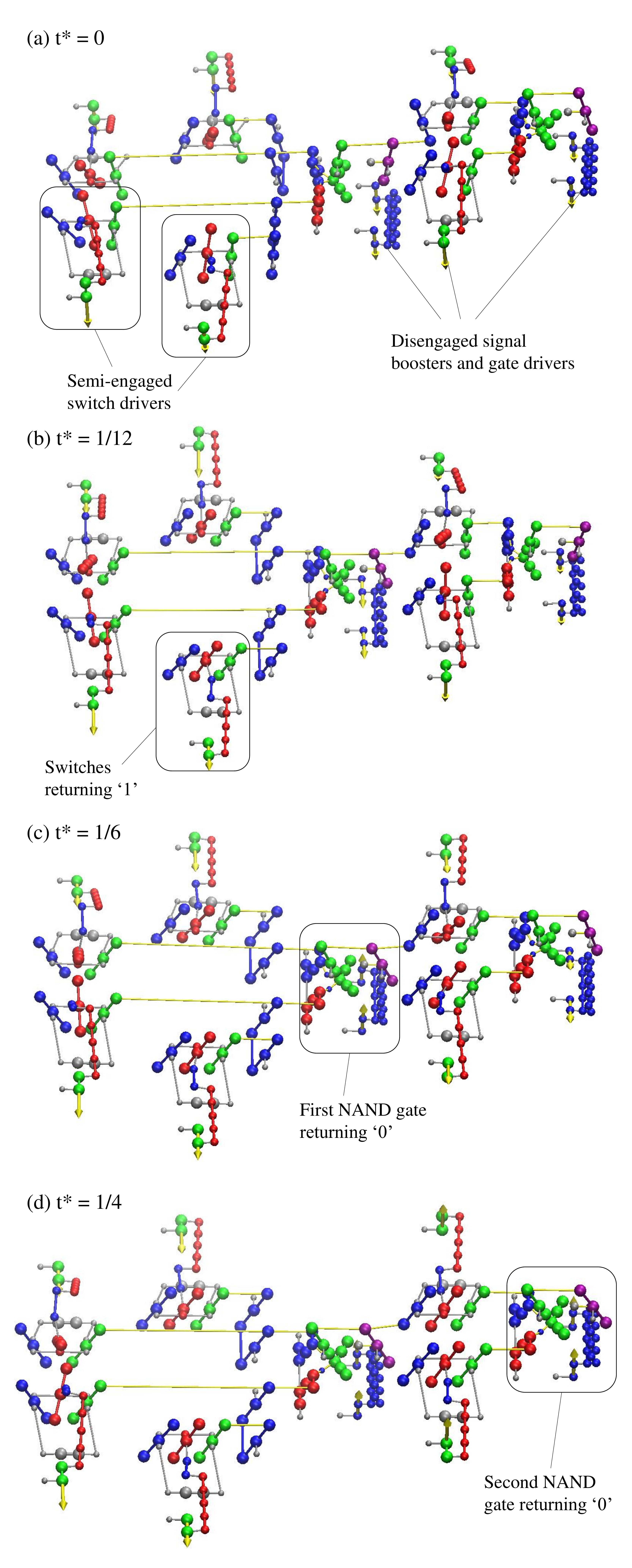}
	\caption{The two-gate phase-chained system at four critical values of $t^*$.}
	\label{fig:phaseChainGallery}
\end{figure}

\begin{table}
\caption{\label{tab:phaseChain}Results obtained from error-rate simulations of four phase-chained NAND gates at $6T_{r}$ (1788\,K) and 2.87\,GHz. No errors were detected over $60\,000$ simulations at $T \leq 3T_r$.}
\begin{ruledtabular}
\begin{tabular}{ccc}
Gate number & 1\textsuperscript{st} phase error & 2\textsuperscript{nd} phase error \\
\midrule
\centering
1 & $5.90\times 10^{-3}$ & $6.00\times 10^{-4}$ \\
2 & $1.00\times 10^{-3}$ & $1.40\times 10^{-3}$ \\
3 & $6.67\times 10^{-4}$ & $5.20\times 10^{-3}$ \\
4 & $5.67\times 10^{-4}$ & $5.43\times 10^{-3}$
\end{tabular}
\end{ruledtabular}
\end{table}

\subsection{\label{ssec:hadder}The Half-Adder}
\begin{figure}
	\includegraphics[width=\linewidth]{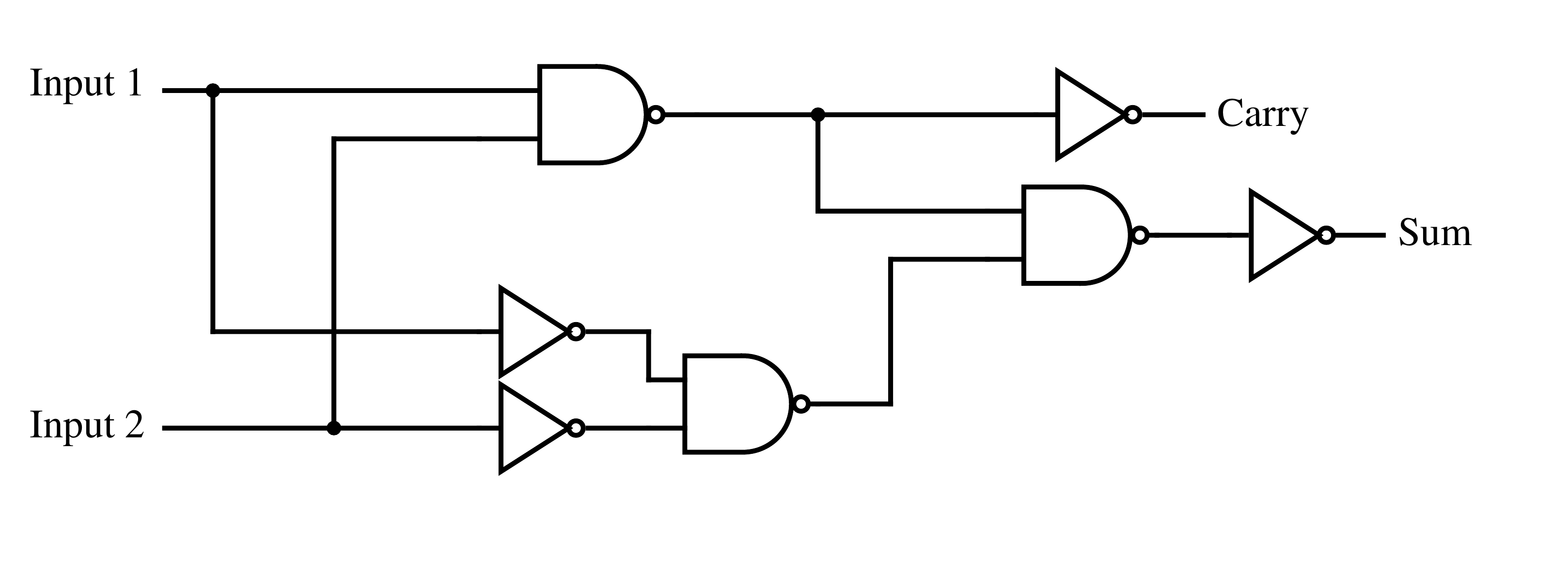}
	\caption{The circuit layout of the half-adder.}
	\label{fig:hadderCircuit}
\end{figure}

To demonstrate the ability of phased chaining to transmit information through complex combinatorial circuits, a half-adder circuit consisting of three phase-chained NAND gates was designed. The half-adder is a combinatorial circuit which receives two binary inputs and returns the sum and carry of the inputs. The sum is equivalent to the Boolean XOR operating on the inputs, and the carry equivalent to the Boolean AND operation performed on the inputs.  Two half-adders may be chained together to create a full-adder, a gate which returns the carry and sum of three different input bits; in turn, a combinatorial gate that can compute the sum of two numbers of arbitrary length can be constructed from full adders alone.  The circuit was constructed from three NAND gates as depicted in Fig.\ \ref{fig:hadderCircuit}.

\begin{figure*}
	\includegraphics[width=0.71\linewidth]{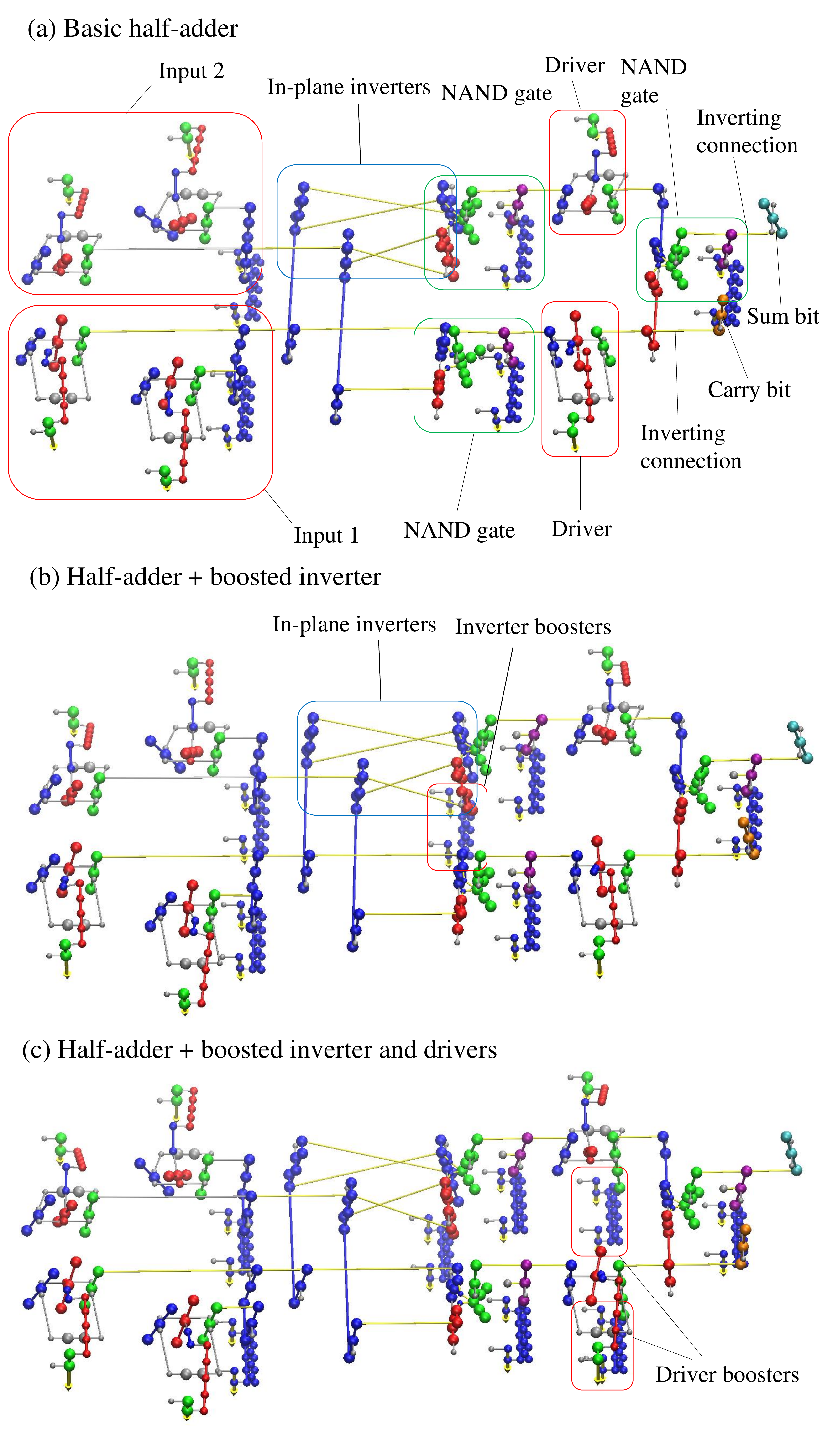}
	\caption{ The basic half-adder and two modifications of the basic design at $t^* = 0$.}
	\label{fig:hadder}
\end{figure*}

\begin{figure}
	\includegraphics[width=\linewidth]{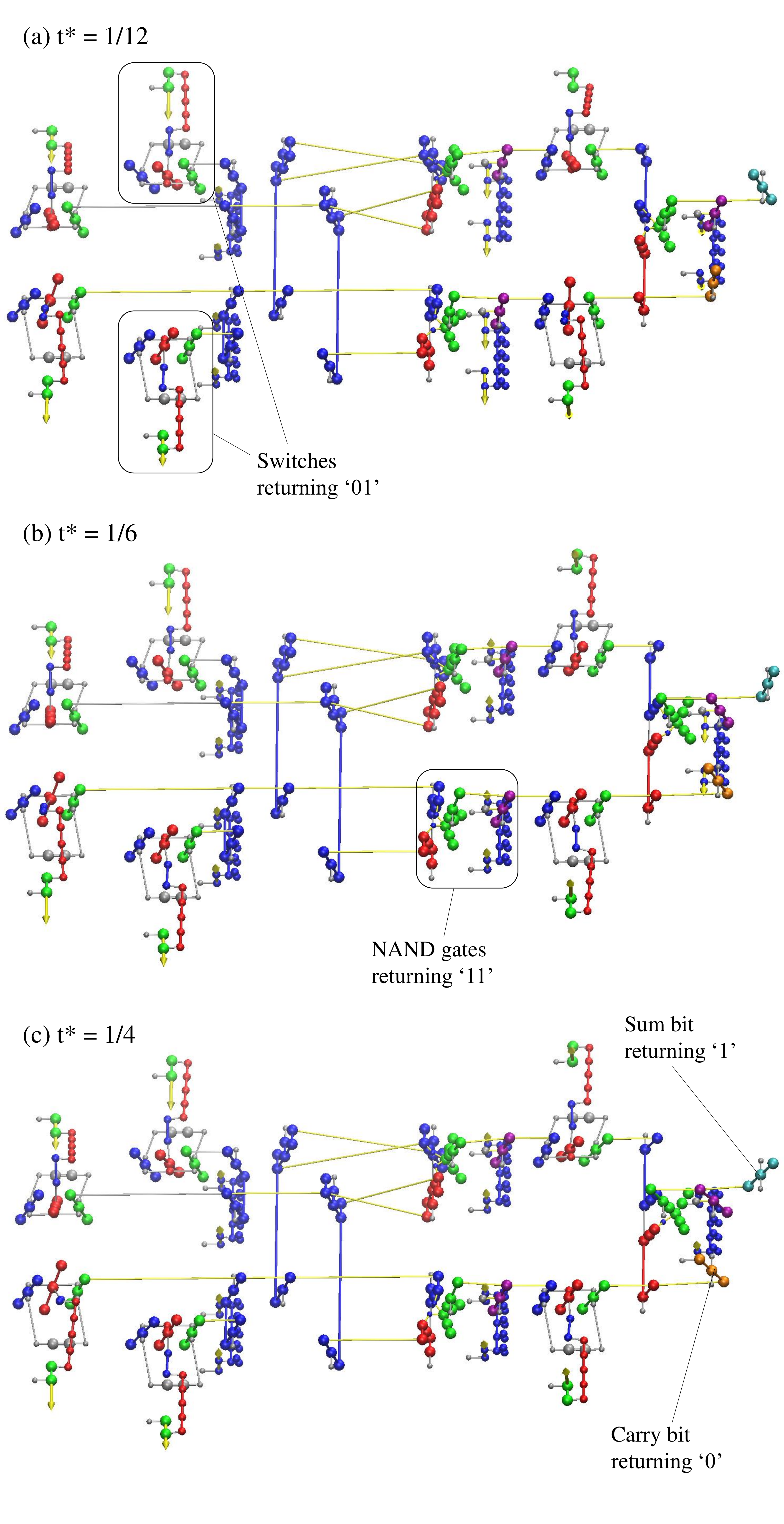}
	\caption{The basic half-adder at three critical values of $t^*$.}
	\label{fig:hadderGallery}
\end{figure}

Operation of the half-adder via the phased chaining protocol proceeds in a manner similar to the chained NAND gates in Section \ref{ssec:phaseChain}. As with the NAND gates, the system is initialised with the switch in the midpoint position and every gate and driver maximally disengaged at $t^* = 0$ (Fig.\ \ref{fig:hadder}(a)). In this particular image, the switches will transition from `01' in the first phase to `11' in the second phase. Thus, the half-adder will return a sum value of `1' and carry value of `0' in the first phase, transitioning to a sum value of `0' and a carry value of `1' in the second phase. At $t^* = 1/12$ (Fig.\ \ref{fig:hadderGallery}(a)), the switches now return `01' and the switch is halted. It should be noted that the design of the switch differs slightly from the design used in the previous section as a booster has been added to increase the strength of the switch output. This modification is necessary as each switch output is coupled to two separate inputs, which increases the error rate thus necessitating an additional booster to mitigate this problem. Information from the switch outputs then passes through the inverting and non-inverting connections to the NAND gates. The upper NAND gate, receiving inverted inputs, is functionally identical to an OR gate.

For this particular set of inputs, both NAND gates will return `1' at $t^* = 1/6$ with their signal boosters fully engaged. Information is then transmitted to the final NAND gate via the drivers, with the signal booster of this gate becoming fully engaged at $t^* = 1/4$. It is at this point that the first-phase error rate of both the cyan-coloured sum bit and orange-coloured carry bit (labelled as `Sum 1' and `Carry 1' respectively in Tables \ref{tab:hadder} to \ref{tab:hadderBoost2}) is measured; for this particular example the sum is `1' and the carry `0'. Between $t^* = 1/4$ and $t^* = 1/2$, this process is reversed, returning the system to its initial state. The second phase is initiated at this point. It is identical to the first phase except that the switches are rotated in the opposite direction such that they return `11'. The second-phase error of both sum and carry bits (labelled as `Sum 2' and `Carry 2' respectively) is therefore measured at $t^* = 3/4$; in this example the second-phase sum is `0' while the carry is `1'.   Supplemental Video 10 provides an animation of this complete process.

In order to simplify the physical construction of the circuit, the in-plane inverter from Section \ref{ssec:circon} was used in place of the simpler inverting connection. However, adopting this type of inverter results in a higher error rate than the directly-tethered connection, as can be observed with the results of the error-rate simulations in Table \ref{tab:hadder}, which are much higher than the results for the NAND gate in Table \ref{tab:muxNand}. Due to the lack of symmetry, there are twelve possible transitions which must be considered for the half-adder.  It is for this reason that a full adder was not simulated, as there are 56 (8 $\times$ 7) possible transitions for a full adder with three possible input bits; therefore, a large increase in computational resources would be required to simulate this gate. 

In order to mitigate the problems posed by the use of the in-plane inverter, an additional booster was placed under the inputs of the topmost NAND gate (Fig.\ \ref{fig:hadder}(b)); this modification serves to effectively boost the inverter outputs. The external dipoles which drive this signal booster have a 90\textdegree\ lag in phase relative to the switch. This, in turn forces the phases of all external dipoles downstream from the inverters to be shifted back by 90\textdegree. Thus, it takes 20\% longer for information to propagate from the switches to the sum and carry bits in this system. In exchange for this reduction in speed, there is a sharp decrease in error rate as can be observed from the results in Table \ref{tab:hadderBoost}.

Further improvements in error rate can be obtained by adding more signal boosters to the drivers which separate the first two NAND gates from the final NAND gate and outputs (Fig.\ \ref{fig:hadder}c). As before, addition of these signal boosters results in a 17\% increase in the time needed for information to propagate along the length of the circuit; this is compensated for by a further large reduction in error rate such that there were no longer any significant errors detected over 4800 simulations at 596\,K. 
 An animation of this half-adder can be viewed in Supplemental Video 11. 
Comparison of the inverter-boosted system at 894\,K (Table \ref{tab:hadderBoostHot}) and the inverter and driver-boosted system at the same temperature (Table \ref{tab:hadderBoost2}) demonstrates the large reduction in error rate.

\begin{figure}
	\includegraphics[width=\linewidth]{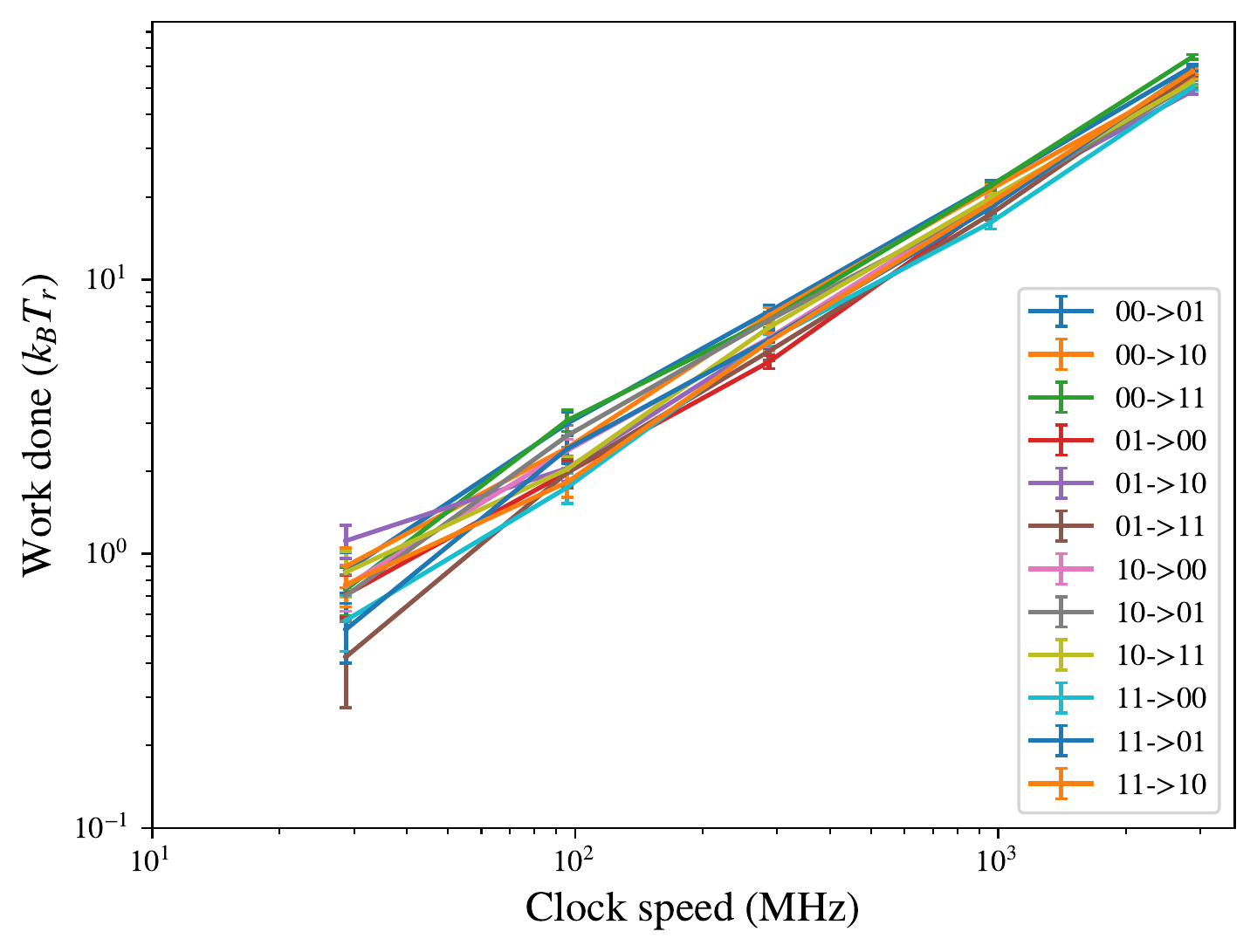}
	\caption{Plot of work done against clock speed in the inverter and driver-boosted half-adder system. Each data point corresponds to an average over 96 simulations.}
	\label{fig:hadderGraph1}
\end{figure}

Thermodynamic reversibility simulations were carried out on the inverter and driver-boosted half-adder using a similar protocol to the chained NAND gates in Section \ref{ssec:phaseChain}. The results are summarised in Fig. \ref{fig:hadderGraph1} and show that the half-adder is tending towards thermodynamic reversibility within the limit of clock speeds tested.
\begin{table}
\caption{\label{tab:hadder}Results obtained from error-rate simulations of the basic half-adder gate performed at $2T_{r}$ (596\,K) for the evolution from a defined starting state to a given target state at a clock speed of 2.87\,GHz. 4800 simulations were carried out for each transition. No errors were detected over $4800$ simulations at $T \leq T_{r}$.}
\begin{ruledtabular}
\begin{tabular}{ccccc}
Transition & Carry 1 & Carry 2 & Sum 1 & Sum 2 \\
\midrule
\centering
00$\rightarrow{}$01 & 0 & 0 & 0 & 0 \\ 
00$\rightarrow{}$10 & 0 & 0& $2.08 \times 10^{-4}$ & 0 \\ 
00$\rightarrow{}$11 & 0& $2.08 \times 10^{-3}$& $6.25 \times 10^{-4}$ & 0 \\ 
01$\rightarrow{}$00 & 0 & 0& $2.08 \times 10^{-4}$& $4.17 \times 10^{-4}$ \\ 
01$\rightarrow{}$10 & 0 & 0 & 0& $2.08 \times 10^{-4}$ \\ 
01$\rightarrow{}$11 & 0& $1.67 \times 10^{-3}$ & 0 & 0 \\ 
10$\rightarrow{}$00 & 0 & 0& $2.08 \times 10^{-4}$ & 0 \\ 
10$\rightarrow{}$01 & 0& $2.08 \times 10^{-4}$& $2.08 \times 10^{-4}$ & 0 \\ 
10$\rightarrow{}$11 & 0& $2.29 \times 10^{-3}$ & 0& $2.08 \times 10^{-4}$ \\ 
11$\rightarrow{}$00& $2.92 \times 10^{-3}$ & 0 & 0& $2.08 \times 10^{-4}$ \\ 
11$\rightarrow{}$01& $6.25 \times 10^{-4}$& $2.08 \times 10^{-4}$ & 0 & 0 \\ 
11$\rightarrow{}$10& $1.67 \times 10^{-3}$ & 0 & 0 & 0 \\ 
\end{tabular}
\end{ruledtabular}
\end{table}

\begin{table}
\caption{\label{tab:hadderBoost}Results obtained from error-rate simulations of the half-adder gate with boosted inverter outputs performed at $2T_{r}$ (596\,K) at a clock speed of 2.87\,GHz. 4800 simulations were carried out for each transition. No errors were detected over $4800$ simulations at $T \leq T_{r}$.}
\begin{ruledtabular}
\begin{tabular}{ccccc}
Transition & Carry 1 & Carry 2 & Sum 1 & Sum 2 \\
\midrule
\centering
00$\rightarrow{}$01 & 0 & 0 & 0 & 0 \\ 
00$\rightarrow{}$10 & 0 & 0 & 0 & 0 \\ 
00$\rightarrow{}$11 & 0& $4.17 \times 10^{-4}$ & 0 & 0 \\ 
01$\rightarrow{}$00 & 0 & 0 & 0 & 0 \\ 
01$\rightarrow{}$10 & 0 & 0 & 0& $2.08 \times 10^{-4}$ \\ 
01$\rightarrow{}$11 & 0& $6.25 \times 10^{-4}$ & 0 & 0 \\ 
10$\rightarrow{}$00 & 0 & 0 & 0 & 0 \\ 
10$\rightarrow{}$01 & 0 & 0 & 0 & 0 \\ 
10$\rightarrow{}$11 & 0& $6.25 \times 10^{-4}$ & 0 & 0 \\ 
11$\rightarrow{}$00& $4.17 \times 10^{-4}$ & 0 & 0 & 0 \\ 
11$\rightarrow{}$01& $6.25 \times 10^{-4}$ & 0 & 0 & 0 \\ 
11$\rightarrow{}$10& $4.17 \times 10^{-4}$ & 0 & 0 & 0 \\ 
\end{tabular}
\end{ruledtabular}
\end{table}

\begin{table}
\caption{\label{tab:hadderBoostHot}Results obtained from error-rate simulations of the half-adder with boosted inverter outputs performed at $3T_{r}$ (894\,K) at a clock speed of 2.87\,GHz. 4800 simulations were carried out for each transition.}
\begin{ruledtabular}
\begin{tabular}{ccccc}
Transition & Carry 1 & Carry 2 & Sum 1 & Sum 2 \\
\midrule
\centering
00$\rightarrow{}$01 & 0& $2.08 \times 10^{-4}$& $2.08 \times 10^{-4}$& $8.33 \times 10^{-4}$ \\ 
00$\rightarrow{}$10 & 0& $6.25 \times 10^{-4}$& $4.17 \times 10^{-4}$& $2.08 \times 10^{-4}$ \\ 
00$\rightarrow{}$11 & 0& $3.96 \times 10^{-3}$& $4.17 \times 10^{-4}$ & 0 \\ 
01$\rightarrow{}$00& $4.17 \times 10^{-4}$ & 0& $2.08 \times 10^{-4}$& $6.25 \times 10^{-4}$ \\ 
01$\rightarrow{}$10& $6.25 \times 10^{-4}$ & 0& $1.04 \times 10^{-3}$& $8.33 \times 10^{-4}$ \\ 
01$\rightarrow{}$11& $1.25 \times 10^{-3}$& $3.13 \times 10^{-3}$& $1.46 \times 10^{-3}$ & 0 \\ 
10$\rightarrow{}$00& $2.08 \times 10^{-4}$ & 0& $6.25 \times 10^{-4}$& $2.08 \times 10^{-4}$ \\ 
10$\rightarrow{}$01& $2.08 \times 10^{-4}$& $6.25 \times 10^{-4}$& $6.25 \times 10^{-4}$& $1.25 \times 10^{-3}$ \\ 
10$\rightarrow{}$11 & 0& $4.38 \times 10^{-3}$& $1.04 \times 10^{-3}$ & 0 \\ 
11$\rightarrow{}$00& $2.92 \times 10^{-3}$ & 0 & 0& $4.17 \times 10^{-4}$ \\ 
11$\rightarrow{}$01& $3.96 \times 10^{-3}$& $1.25 \times 10^{-3}$& $2.08 \times 10^{-4}$& $1.25 \times 10^{-3}$ \\ 
11$\rightarrow{}$10& $4.38 \times 10^{-3}$& $4.17 \times 10^{-4}$ & 0& $1.04 \times 10^{-3}$ \\ 
\end{tabular}
\end{ruledtabular}
\end{table}

\begin{table}
\caption{\label{tab:hadderBoost2}Results obtained from error-rate simulations of the half-adder with boosted inverter and driver outputs performed at $3T_{r}$ (894\,K) at a clock speed of 2.87\,GHz. 4800 simulations were carried out for each transition. No errors were detected over $4800$ simulations at $T \leq 2T_{r}$.}
\begin{ruledtabular}
\begin{tabular}{ccccc}
Transition & Carry 1 & Carry 2 & Sum 1 & Sum 2 \\
\midrule
\centering
00$\rightarrow{}$01 & 0 & 0& $1.46 \times 10^{-3}$ & 0 \\ 
00$\rightarrow{}$10 & 0 & 0& $1.04 \times 10^{-3}$ & 0 \\ 
00$\rightarrow{}$11 & 0& $6.25 \times 10^{-4}$& $8.33 \times 10^{-4}$& $6.25 \times 10^{-4}$ \\ 
01$\rightarrow{}$00 & 0 & 0 & 0& $2.08 \times 10^{-3}$ \\ 
01$\rightarrow{}$10& $1.25 \times 10^{-3}$ & 0& $1.25 \times 10^{-3}$ & 0 \\ 
01$\rightarrow{}$11& $8.33 \times 10^{-4}$ & 0& $8.33 \times 10^{-4}$ & 0 \\ 
10$\rightarrow{}$00 & 0 & 0 & 0& $4.17 \times 10^{-4}$ \\ 
10$\rightarrow{}$01 & 0& $4.17 \times 10^{-4}$ & 0& $4.17 \times 10^{-4}$ \\ 
10$\rightarrow{}$11& $2.08 \times 10^{-4}$ & 0& $2.08 \times 10^{-4}$ & 0 \\ 
11$\rightarrow{}$00 & 0 & 0 & 0 & 0 \\ 
11$\rightarrow{}$01 & 0 & 0& $6.25 \times 10^{-4}$& $6.25 \times 10^{-4}$ \\ 
11$\rightarrow{}$10& $4.17 \times 10^{-4}$ & 0& $4.17 \times 10^{-4}$ & 0 \\ 
\end{tabular}
\end{ruledtabular}
\end{table}

\section{Conclusion}
Despite being subject to a large number of design constraints, molecular mechanical logic gates potentially allow for the design of compact reversible logical systems capable of functioning efficiently at room temperature and under standard atmospheric conditions. We have presented a general framework for constructing thermodynamically reversible combinatorial circuits, which allows for the investigation of thermodynamic phenomena in logical systems that are not purely abstract. We have demonstrated Landauer's principle and an efficient novel input-preserving molecular-mechanical NAND gate design. Although the experimental synthesis of a molecular-scale system based on this framework is likely to remain impractical in the near future, our work nevertheless demonstrates the theoretical soundness of using molecular mechanical logic gates to construct reversible circuits within this framework.

Using the aforementioned framework, we have demonstrated the importance of staggering the phases of chained gates when comparing directly-chained and phase-chained protocols, and the use of signal boosting to reduce the error rate in more complex systems such as the half-adder. In comparing the performance of the boosted and unboosted systems, we find a trade-off between the complexity of the control protocol that drives a computation, and the accuracy of that computation. Such a trade-off is only visible when a concrete - albeit idealised - computational architecture is modelled.

In addition to the work we have already performed, our framework also presents many possibilities for future work and expansion. One such possibility is the investigation of systems which rotate continuously and can thus be driven by a single external dipole. These systems offer the possibility of greater speed and reduced number of external inputs at the cost of increased error rate and the possibility of logical irreversibility in the event that intermediate inputs are not preserved. A further possibility involves extending the framework to allow for sequential logic, via a modification of the double-bit system. The inclusion of registers or other such methods of persistent information storage would in theory allow for the construction of a Turing-complete device. As with driving the system via a single external dipole, such an extension can result in potential logical irreversibility due to intermediate bits being overwritten during the operation of the device. Further, the added complexity would greatly increase the difficulty of human optimisation of the circuit; overcoming this barrier may be possible via optimisation with a genetic algorithm or other meta-heuristic.  Finally, the sophistication of the simulation could be increased by decomposing the rigid bodies into point masses connected by stiff bonding and angular potentials, though this fine-graining would result in an increase in the computational resources required to simulate a given system. Introducing flexibility into the rigid bodies that currently compose the system may result in an increase in error rate, although it is possible thermodynamic efficiency might improve as the components also have more flexibility to avoid impinging on each other. 

\section{Additional Materials}
VMD-viewable trajectory files of the processes illustrated in the Supplemental Videos are available at {the Oxford University Research Archive:} \url{https://ora.ox.ac.uk/objects/uuid:978da7b2-de5b-49c7-ac39-69fabfbc8182}.
\\

\begin{acknowledgments}
The authors are grateful to the UK Materials and Molecular Modelling Hub and to the University of Oxford Advanced Research Computing (ARC) facility for computational resources. T.E. Ouldridge is supported by a Royal Society University Research Fellowship.
\end{acknowledgments}

\bibliography{apssamp}% Produces the bibliography via BibTeX.

\end{document}